\documentclass[12pt,letterpaper]{article}
\usepackage[dvips]{graphicx,color}
\usepackage{amsmath,amssymb}
\usepackage[dvips,letterpaper,text={6.5in,9in}]{geometry}
\usepackage{fancyhdr}
\usepackage{verbatim}

\newcommand\ltap{\
  \raise.3ex\hbox{$<$\kern-.75em\lower1ex\hbox{$\sim$}}\ } 
\newcommand\gtap{\
  \raise.3ex\hbox{$>$\kern-.75em\lower1ex\hbox{$\sim$}}\ } 

\newcommand\simge{\mathrel{%
   \rlap{\raise 0.511ex \hbox{$>$}}{\lower 0.511ex \hbox{$\sim$}}}}
\newcommand\simle{\mathrel{%
   \rlap{\raise 0.511ex \hbox{$<$}}{\lower 0.511ex \hbox{$\sim$}}}}

\newcommand{\slashchar}[1]%
        {\kern .25em\raise.18ex\hbox{$/$}\kern-.75em #1}
\def\lsim{\mathrel{\raise.3ex\hbox{$<$\kern-.75em\lower1ex\hbox{$\sim$}}}}
\def\gsim{\mathrel{\raise.3ex\hbox{$>$\kern-.75em\lower1ex\hbox{$\sim$}}}}

\newcommand\CD{{\cal D}}

\newcommand\CH{{\cal H}}

\newcommand\CM{{\cal M}}

\newcommand\CO{{\cal O}}

\newcommand\CU{{\cal U}}

\newcommand\CX{{\cal X}}
\newcommand\CY{{\cal Y}}
\newcommand\CZ{{\cal Z}}
\newcommand\be{\begin{equation}} 
\newcommand\ee{\end{equation}} 
\newcommand\bea{\begin{eqnarray}}
\newcommand\eea{\end{eqnarray}}
\newcommand\ba{\begin{array}}
\newcommand\ea{\end{array}}
\newcommand\nn{\nonumber}

\newcommand{\lcp}{\ensuremath{\lambda_{CP}}}

\newcommand{\gm}{\ensuremath{\gamma^{\mu}}}

\newcommand{\half}{\ensuremath{\frac{1}{2}}}
\newcommand{\thhalf}{\ensuremath{\frac{3}{2}}}

\newcommand{\BD}{\ensuremath{B_d}}
\newcommand{\BDbar}{\ensuremath{\bar B_d}}
\newcommand{\stwobeta}{\ensuremath{\sin{2\beta}}}
\newcommand\thc{\theta_C}
\newcommand\thy{\theta_Y}
\newcommand\dagg{\dagger}
\newcommand\ts{\thinspace}
\newcommand\ra{\rightarrow}

\newcommand\ol{\bar}
\newcommand\mev{{\rm MeV}}
\newcommand\gev{{\rm GeV}}
\newcommand\tev{{\rm TeV}}

\newcommand\TeV{{\rm TeV}}

\newcommand\nin{\noindent}
\newcommand\lvac{\langle \Omega \vert}
\newcommand\rvac{\vert \Omega \rangle}

\newcommand\condtbt{\langle \bar t t\rangle}

\newcommand\getc{g_{ETC}}

\newcommand\Gew{SU(2)\otimes U(1)}
\newcommand\Getc{G_{ETC}}

\newcommand\suc{SU(3)_C}
\newcommand\Ntc{N_{TC}}
\newcommand\sutc{SU(N_{TC})}
\newcommand\uone{U(1)_1}
\newcommand\utwo{U(1)_2}

\newcommand\suone{SU(3)_1}
\newcommand\sutwo{SU(3)_2}

\newcommand\atc{\alpha_{TC}}

\newcommand\Metc{M_{ETC}}

\newcommand\Ltc{\Lambda_{TC}}
\newcommand\condtc{{\langle \ol T T \rangle}_{TC}}
\newcommand\condetc{{\langle \ol T T \rangle}_{ETC}}

\begin{document}
\title{
\vskip -15mm
\begin{flushright}
\vskip -15mm
{\small BUHEP-04-03\\
hep-ph/0404107\\}
\vskip 5mm
\end{flushright}
{\Large{\bf CP Violation and Mixing in Technicolor Models}}\\
}
\author{
{\large Adam Martin\thanks{aomartin@bu.edu}\,\,\, and Kenneth
  Lane\thanks{lane@bu.edu}}\\
{\large {$$}Department of Physics, Boston University}\\
{\large 590 Commonwealth Avenue, Boston, Massachusetts 02215}\\
}
\maketitle

\begin{abstract}
  Vacuum alignment in technicolor models provides an attractive origin for
  the quarks' CP violation and, possibly, a natural solution for the
  strong-CP problem of QCD. We discuss these topics in this paper. Then we
  apply them to determine plausible mixing matrices for left and right-handed
  quarks. These matrices determine the Cabibbo-Kobayashi-Maskawa matrix as
  well as new mixing angles and phases that are observable in extended
  technicolor (ETC) and topcolor (TC2) interactions. We determine the
  contributions of these new interactions to CP-violating and mixing
  observables in the $K^0$, $B_d$ and $B_s$ systems. Consistency with mixing
  and CP violation in the $K^0$ system requires assuming that ETC
  interactions are electroweak generation conserving even if technicolor has
  a walking gauge coupling. Large ETC gauge boson masses and small
  intergenerational mixing then result in negligibly small ETC contributions
  to $B$-meson mixing and CP violation and to ${\rm Re}(\epsilon'/\epsilon)$.
  We confirm our earlier strong lower bounds on TC2 gauge boson masses from
  $B_d$--$\ol B_d$ mixing. We then pay special attention to the possibility
  that current experiments indicate a deviation from standard model
  expectations of the values of $\sin 2\beta$ measured in $B_d \ra J/\psi
  K_S$, $\phi K_S$, $\eta' K_S$, and $\pi K_S$, studying the ability of TC2
  to account for these. We also determine the TC2 contribution to $\Delta
  M_{B_s}$ and to ${\rm Re}(\epsilon'/\epsilon)$, and find them to be
  appreciable.
\end{abstract}


\newpage

\section*{1. Introduction and Overview}

In this paper we study predictions of topcolor-assisted technicolor models
for CP violation in the $K^0$ and $B^0$ systems. We are particularly
interested in determining whether these or similar models with CP-violating
flavor-changing neutral currents can account for the apparent discrepancies
with standard model predictions of the parameters measured in $B_d \ra
J/\psi K_S$, $\phi K_S$, $\eta' K_S$ and $\pi K_S$:
\bea\label{eq:XKsData}
\sin{2\beta}_{J/\psi K_S} &=& +0.72\pm
0.05\qquad\qquad\,\,{\cite{Browder:2003ii}}\nn\\
\sin{2\beta}_{\phi K_S} &=& +0.50\pm 0.25\quad({\rm Babar}\,\,
{\cite{Aubert:2004dy}})\nn\\
\sin{2\beta}_{\phi K_S} &=& +0.06\pm 0.33\quad({\rm Belle}\,\,\,\,\,
{\cite{Abe:2004xp}})\\
\sin{2\beta}_{\eta' K_S} &=& +0.27\pm
0.21\,\,\qquad\qquad{\cite{Aubert:2003bq}}\nn\\
\sin{2\beta}_{\pi K_S} &=& +0.48^{+0.38}_{-0.47}\pm0.11
\quad\quad\,\,\,{\cite{Aubert:2004xf}}\nn
\eea

Topcolor-assisted technicolor (TC2) is the most fully-developed {\em
  dynamical} description of electroweak and flavor physics (for recent
reviews, see Refs.~\cite{Hill:2002ap} and~\cite{Lane:2002wv}). It consists of
strong technicolor (TC) and topcolor gauge interactions that induce
spontaneous breakdown of electroweak $\Gew$ symmetry to $U(1)_{EM}$ {\em and}
a large top quark condensate $\condtbt \sim (100\,\gev)^3$ and mass $m_t
\simeq 170\,\gev$. The strong gauge groups, plus color and at least part of
electroweak $U(1)$, are embedded in an extended technicolor (ETC) gauge group
$\Getc$~\cite{Eichten:1980ah} which, when broken at high energies, provides
the $5\,\mev$ to $5\,\gev$ hard masses of all standard model fermions,
including the top quark. The masses of ETC gauge bosons range from $\Metc
\simeq 10$--$50\,\tev$ for $m_q(\Metc) \simeq 5\,\gev$ up to $\Metc \simeq
2000$--$20,000\,\tev$ for $m_q(\Metc) \simeq 5\,\mev$.\footnote{Appendix~A
  contains our estimates of $\Metc/\getc$ in generic TC2 models with walking
  technicolor.} Such large ETC masses for the light quarks are necessary to
adequately suppress their CP-conserving and violating $|\Delta S| = 2$
interactions. Reasonable quark masses are then possible because of the
``walking'' technicolor gauge coupling~\cite{Holdom:1981rm,Appelquist:1986an,
  Yamawaki:1986zg,Akiba:1986rr} that strongly enhances the technifermion
condensate $\condetc$.

In TC2 models, the large top, but not bottom, condensate and mass is due to
$\suone \otimes \uone$ gauge interactions which are strong near
1~TeV~\cite{Hill:1995hp}. The $\suone$ interaction is $t$--$b$ symmetric
while $\uone$ couplings are $t$--$b$ asymmetric.\footnote{Electroweak $SU(2)$
  and $U(1)_{1,2}$ commute.} In particular, the $\uone$ hypercharges of $t$
and $b$ must satisfy $Y_{1Lt} Y_{1Rt} > 0$ and, probably, $Y_{1Lb} Y_{1Rb} <
0$ (see Sect.~4). This makes these forces supercritical for breaking the top
quark chiral symmetry, but subcritical for bottom.\footnote{We assume
  throughout this paper that the effective $\uone$ coupling should be strong,
  at least for the top and bottom quarks, so that the $\suone$ coupling does
  not need to be fine-tuned for top condensation. This raises the concern
  that the $\uone$ coupling has a Landau pole at low
  energy~\cite{Chivukula:1998vd}. One resolution of this difficulty is that
  the embedding of $\uone$ and other TC2 gauge groups into $\Getc$ ---
  necessary to avoid an axion~\cite{Eichten:1980ah} --- occurs at a low
  enough energy to forestall the Landau pole.} There are weaker $\sutwo
\otimes \utwo$ gauge interactions in which light quarks (and leptons) may or
may not participate.

For TC2 to be consistent with precision measurements of the
$Z^0$~\cite{Chivukula:1996cc}, the two $U(1)$'s must be broken to weak
hypercharge $U(1)_Y$ at an energy somewhat higher than 1~TeV by
electroweak-singlet condensates~\cite{Lane:1996ua}. This breaking results in
a heavy color-singlet $Z'$ boson which plays a central role in this paper.
The two $SU(3)$'s are broken at $1\,\tev$ to their diagonal $\suc$ subgroup
--- ordinary color. A massive octet of ``colorons'', $V_8$, mediate the
broken topcolor $SU(3)$ interactions.

There are two variants of TC2: The ``standard'' version, which we denote
STC2~\cite{Hill:1995hp}, in which only the third generation quarks are
$\suone$ triplets. The third generation quarks also transform under $\uone$.
{\em Whether the lighter two quark generations also transform under $\uone$
  is a model-dependent question. In this paper we assume that they do.}
Indeed, in some models, $\uone$ anomaly cancellation may require
it~\cite{Lane:1996ua}.  In STC2 strongly-coupled flavor-changing neutral
current (FCNC) interactions are mediated by both $V_8$ and $Z'$
exchange.\footnote{Of course, ETC interactions induce FCNC effects as well.}
In order that they not be prohibitively large for the light quarks, the first
two generations must have flavor-symmetric $\uone$ hypercharges, i.e., for
the electroweak eigenstates,
\bea\label{eq:YGIM}
&& Y_{1Lu} = Y_{1Ld} = Y_{1Lc} = Y_{1Ls} \,,\nn\\
&& Y_{1Ru} = Y_{1Rc}, \quad Y_{1Rd} = Y_{1Rs}\,.
\eea

The other variant is the ``flavor-universal'' version,
FUTC2~\cite{Chivukula:1996yr,Popovic:1998vb}.  There, all quarks are $\suone$
triplets. The third generation quarks transform under $\uone$ and we assume
again that the light generations do too. In FUTC2, only $Z'$ exchange induces
new FCNC interactions. Therefore, the $\uone$ hypercharges of light quarks
must satisfy Eq.~(\ref{eq:YGIM}) here as well. We consider both TC2 variants
in this paper.

In Sect.~2 we review vacuum alignment in technicolor theories and show how
this determines CP violation in quark interactions. Vacuum alignment is the
process in which ETC and TC2 interactions lift the degeneracy of the infinity
of vacua associated with spontaneous breaking of technifermion and quark
chiral symmetries. It can induce CP violation in a theory which is
superficially CP invariant. This leads to a new, natural scenario for solving
the strong-CP problem of QCD.\footnote{Some of the material presented in
  Sect.~2 appeared in the conference proceedings
  Refs.~\cite{Lane:2001ar,Lane:2001rv,Lane:2002wv}} Alignment generates the
matrices $Q_{L,R} = (U,D)_{L,R}$ that rotate left and right-handed up and
down quarks from the electroweak basis to the mass-eigenstate one. The
Cabibbo-Kobayashi-Maskawa (CKM) matrix is $V = U_L^\dagg D_L$. Observable
CP-violating phases appear in the ordinary weak interactions through $V$ {\em
  and} in the TC2 and ETC interactions through $Q_{L,R}$. We review and
update general constraints on the form of the alignment matrices in Sect.~2.

In Sect.~3 we describe the TC2 and ETC interactions we use for $B^0$ and
$K^0$ studies in later sections. Here we show how the quark alignment
matrices enter these interactions. We also discuss an important assumption we
must make for ETC interactions, namely, that they are electroweak generation
conserving. In Sect.~4 we discuss the main constraints on ETC and TC2 that
arise from the requirement that TC2 causes no quark other than the top to
condense, and from neutral meson mixing and CP violation. Mixing of $\BD$ and
$\ol \BD$ leads to lower bounds on $M_{V_8}$ and $M_{Z'}$. We confirm the
bounds found earlier in Ref.~\cite{Burdman:2000in}, and we are in some
disagreement with a later study by Simmons~\cite{Simmons:2001va}.  In Sect.~5
we present the formalism for calculating the TC2 contributions to $B_d \ra X
K_S$. Because of our assumption of electroweak generation conservation, the
ETC contributions are negligible.  Section~6 is a brief review of the
definition of the {\em experimental} $\sin 2\beta_{\rm eff}$, as opposed to
the standard model value, $(\sin 2\beta)_{SM} = \sin (2 \arg(V^*_{td}))$,
where $V_{td}$ is the CKM matrix element.

Section~7 contains our main results. They are based on three ``models'' of
the quark mass matrix that are inspired by the CP-violation scenario
described in Sect.~2. Once the mass matrices are written down, the alignment
matrices $Q_{L,R}$ are determined. We use these to compare the predictions of
TC2 with experiment for $\BD \ra J/\psi K_S$, $\phi K_S$, $\eta' K_S$ and
$\pi K_S$.  We also calculate the influence of TC2 on $x_s = \Delta
M_{B_s}/\Gamma_{B_s}$ and ${\rm Re}(\epsilon'/\epsilon)$.  Our main
conclusions are these: (1) If $D_R$ is $2\times 2$ by $1\times 1$
block-diagonal --- as {\em may} be necessary to avoid excessive $\BD$--$\ol
\BD$ mixing, both TC2 variants predict that the same value of $\sin 2\beta$
is measured in all these processes and that this value is the one expected in
the standard model --- even though TC2 may contribute appreciably to the
decay amplitudes. (2) If $D_R$ is not block-diagonal, the value of $\sin
2\beta$ extracted from $B_d \ra J/\psi K_S$ is the standard model
expectation, but other $B_d \ra X K_S$ decays may lead to different values
and even fit the central values of current measurements.\footnote{Burdman has
  recently carried out similar studies ~\cite{Burdman:2003nt} in which he
  considered the effects of warped extra dimensions and of topcolor $V_8$
  (but not $Z'$) exchange.} We find that TC2 can account for discrepancies as
large as those in the central values of, say, the current Belle measurement,
but this typically requires large $\uone$ hypercharges, especially in the
FUTC2 variant. These are worrisome because they suggest the $\uone$ coupling
has a Landau pole at relatively low energies~\cite{Chivukula:1998vd}. This
problem is less pronounced with STC2 than with FUTC2 because the latter
variant has only the $Z'$ to influence the decays. (3) Depending on the
mixing angles in the $D_{L,R}$ alignment matrices, we find that TC2 (plus the
standard model) can produce values of $x_s$ ranging from the experimental
lower bound of about 20 up to several hundred. (4) For the mass matrix models
we considered, the standard model contributions to ${\rm
  Re}(\epsilon'/\epsilon)$ happened to be about 1/4 the measured value. With
$Z'$ and $V_8$ masses consistent with $\BD$--$\ol \BD$ mixing, the TC2
contribution then can account for the remainder so long as we require $Y_{1Ru}
\cong Y_{1Rd}$.

\section*{2. Vacuum Alignment and CP Violation in Technicolor}

Quark CP violation in technicolor models is a consequence of ``vacuum
alignment''~\cite{Dashen:1971et,Eichten:1980du, Lane:1981je}. The idea is
simple: In technicolor, large flavor/chiral symmetries of the technifermions
and quarks are spontaneously broken when their dynamics become strong. This
leads to an infinity of degenerate vacua. The degeneracy is at least partly
lifted by ETC interactions which explicitly break all global flavor
symmetries.  Vacuum alignment is the process of finding the (perturbative)
ground state which minimizes the expectation value of the chiral-symmetry
breaking Hamiltonian $\CH'$. This Hamiltonian is generated by the exchange of
ETC gauge bosons and it is natural to assume that it is CP-conserving. As
Dashen first showed, however, the ground state $\rvac$ which minimizes
$\lvac\CH'\rvac$ may not respect the same CP symmetry that $\CH'$ does. In
this case, CP is spontaneously broken and $\rvac$ is discretely
degenerate.\footnote{We are aware that spontaneous CP violation at 1~TeV
  implies a significant domain-wall problem. Should this mechanism prove
  successful, we are confident that cosmologists will find a way to eliminate
  the problem.} As we discuss below, this scenario offers the possibility of
naturally solving the strong-CP problem of QCD --- without an axion and
without a massless up quark.

New CP-violating phases are introduced into the $K^0$ and $B^0$ decay
amplitudes by quark-alignment matrices $Q_{L,R}$. To understand how the
$Q_{L,R}$ are determined, we briefly describe vacuum alignment in technicolor
theories. Readers familiar with this material can skip to
Eq.~(\ref{eq:primordial}).

As in Ref.~\cite{Lane:2000es}, we consider simple models in which a single
kind of technifermion interacts with itself and with quarks via ETC
interactions. Leptons are ignored. There are $N$ doublets of these
technifermions, $T_{L,R\, I} = (\CU_{L,R\, I}, \, \CD_{L,R\, I})$, $I =
1,2,\dots,N$, all assumed to transform according to the fundamental
representation of the TC gauge group $\sutc$.  They are ordinary
color-singlets.\footnote{This is not correct for TC2 where some
  technifermions are expected to be triplets under $\suone$ or $\sutwo$.
  This complication is not important for the analysis of $K^0$ and $B^0$
  decays in later sections.} There are three generations of $SU(3)_C$ triplet
quarks $q_{L,R\, i} = (u_{L,R\, i}, \, d_{L,R\, i})$, $i = 1,2,3$.
Left-handed fermions are electroweak $SU(2)$ doublets and right-handed ones
are singlets. Here and below, we exhibit only flavor, not TC and QCD,
indices.

The joint $T$--$q$ chiral flavor group of our model is $G_f =
\left[SU(2N)_{L} \otimes SU(2N)_{R}\right]_T \otimes\left[SU(6)_{L} \otimes
  SU(6)_{R}\right]_q$.\footnote{The fact that heavy quark chiral symmetries
  cannot be treated by chiral perturbative methods will be addressed below.
  We have excluded anomalous $U_A(1)$'s strongly broken by TC and color
  instanton effects. Therefore, alignment matrices must be unimodular.} When
the TC and QCD couplings reach critical values, these symmetries are
spontaneously broken to $S_f = SU(2N) \otimes SU(6)$. Rather than fix the
symmetry-breaking Hamiltonian and vary over the ground states, it is
convenient to work in a ``standard vacuum'' $|\Omega\rangle$ whose symmetry
group is the the vectorial $SU(2N)_V \otimes SU(6)_V$, and chirally rotate
$\CH'$.  Fermion bilinear condensates in $\rvac$ have the simple form
\bea\label{eq:standard}
\langle \Omega |\ol \CU_{LI} \CU_{RJ}|\Omega \rangle &=&
\langle \Omega |\ol \CD_{LI} \CD_{RJ}|\Omega \rangle = -\delta_{IJ} \Delta_T
\nn\\
\langle \Omega |\ol u_{Li} u_{Rj}|\Omega \rangle &=&
\langle \Omega |\ol d_{Li} d_{Rj}|\Omega \rangle = -\delta_{ij} \Delta_q \,.
\eea
Here, $\Delta_T \simeq 2\pi F_T^3$ and $\Delta_q \simeq 2\pi f_\pi^3$ where
$F_T = 246\,\gev/\sqrt{N}$ is the technipion decay constant.\footnote{In TC2
  models with topcolor breaking by technifermion
  condensation~\cite{Lane:1996ua}, $N \sim 10$. This large $N$ raises the
  question of technicolor's contribution to precisely measured electroweak
  quantities such as $S$, $T$, and $U$. Calculations that show technicolor to
  be in conflict with precision measurements have been based on the
  assumption that technicolor dynamics are just a scaled-up version of
  QCD~\cite{Peskin:1990zt,Holdom:1990tc,Golden:1991ig}. However, this cannot
  be because of the walking TC gauge coupling~\cite{Lane:2002wv}. In walking
  technicolor there must be something like a tower of spin-one technihadrons
  reaching almost to the ETC scale, and these states contribute significantly
  to the integrals over spectral functions involved in calculating $S$, $T$,
  and $U$. Therefore, in the absence of detailed experimental knowledge of
  this spectrum, including the spacing between states and their coupling to
  the electroweak currents, it is not possible to calculate $S$, $T$, $U$
  reliably.}

We write the ETC Hamiltonian in the phenomenological four-fermion form (sum
over repeated flavor indices)~\footnote{We assume that ETC interactions
  commute with electroweak $SU(2)$, though not with $U(1)$ nor color $SU(3)$.
  All fields in Eq.~(\ref{eq:Hetc}) are electroweak, not mass, eigenstates.
  In writing $\CH'$, we assume that topcolor breaking to $\suc\otimes U(1)_Y$
  has occurred. Broken topcolor interactions can always be put in the form of
  terms appearing in $\CH'$.}
\bea\label{eq:Hetc}
\CH' &\equiv& \CH'_{TT} + \CH'_{Tq} + \CH'_{qq} \nn\\
&=& \Lambda^{TT}_{IJKL} \, \ol{T}_{LI}\gamma^{\mu}T_{LJ}
\, \ol{T}_{RK}\gamma_{\mu}T_{RL} \nn 
+ \Lambda^{Tq}_{IijJ} \, \ol{T}_{LI}\gamma^{\mu}q_{Li}
\, \ol{q}_{Rj}\gamma_{\mu}T_{RJ} + {\rm h.c.}\\
&+& \Lambda^{qq}_{ijkl} \, \ol{q}_{Li}\gamma_{\mu}q_{Lj}
\, \ol{q}_{Rk}\gamma_{\mu}q_{Rl} + {\rm LL~~and~~RR~~terms.}
\eea
Here, the fields $T_{L,R\, I}$ and $q_{L,R\, i}$ stand for all $2N$
technifermions and six quarks, respectively. The $\Lambda$ coefficients are
$\CO(g^2_{ETC}/M^2_{ETC})$ times ETC boson mixing factors and group
theoretical factors for the broken generators of ETC. The LL and RR terms do
not enter vacuum alignment, but they can be important for quark FCNC
interactions. The $\Lambda$'s may have either sign. In all calculations, we
must choose the $\Lambda$'s to avoid very light pseudoGoldstone bosons (e.g.,
axions). Hermiticity of $\CH'$ requires $(\Lambda^{TT}_{IJKL})^* =
\Lambda^{TT}_{JILK}$, etc.
%
%
Assuming, for simplicity, that color and technicolor are embedded in a simple
nonabelian ETC group, the instanton angles $\theta_{TC}$ and $\theta_{QCD}$
are equal. Without loss of generality, we may work in vacua in which they are
zero. Then strong-CP violation in QCD is characterized by $\ol\theta_q =
\arg\det(M_q)$, where $M_q$ is the quark mass matrix. The assumption of
time-reversal invariance for this theory before any potential breaking via
vacuum alignment then means that all the $\Lambda$'s are {\em real} and so
$\Lambda^{TT}_{IJKL} = \Lambda^{TT}_{JILK}$, etc.

Vacuum alignment now proceeds by minimizing the expectation value of $\CH'$
rotated by elements of $G_f/S_f$. Make the transformations $T_{L,R} \ra
W_{L,R} \, T_{L,R}$ and $q_{L,R} \ra Q_{L,R} \, q_{L,R}$, where $W_{L,R}
\in SU(2N)_{L,R}$ and $Q_{L,R} \in SU(6)_{L,R}$. Then
\bea\label{eq:HW}
\CH'(W,Q) &=& \CH'_{TT}(W_L,W_R) +  \CH'_{Tq}(W,Q) +
\CH'_{qq}(Q_L,Q_R) \\
&=& \Lambda^{TT}_{IJKL} \, \ol{T}_{LI'} W_{L\, I'I}^\dag
\gamma^{\mu}W_{L\, JJ'}T_{LJ'} \, \ol{T}_{RK'} W_{R\, K'K}^\dag
\gamma^{\mu}W_{R\, LL'}T_{RL'} + \cdots \,.\nn
\eea

Since $T$ and $q$ transform according to complex representations of their
respective color groups, the vacuum energy to be minimized has the form
\bea\label{eq:vacE}
& &E(W,Q) \equiv \lvac \CH'(W,Q) \rvac
= E_{TT}(W) + E_{Tq}(W,Q) + E_{qq}(Q) \\
& & \,\, = -\Lambda^{TT}_{IJKL} \, W_{JK} \, W^\dag_{LI} \, \Delta_{TT}
       -\left(\Lambda^{Tq}_{IijJ} \, Q_{ij} \, W^\dag_{JI} + {\rm c.c.}
         \right) \Delta_{Tq} 
       -\Lambda^{qq}_{ijkl} \, Q_{jk} \, Q^\dag_{li} \, \Delta_{qq} \nn \\
& & \,\, = -\Lambda^{TT}_{IJKL} \, W_{JK} \, W^\dag_{LI} \, \Delta_{TT}
\,(1+\rho) \,.\nn 
\eea
The factor $\rho = \CO(10^{-11})$ is explained below. Vacuum alignment must
preserve electric charge conservation, and so the minimum of $E$ occurs in
the subspace of block-diagonal alignment matrices
\bea\label{block}
W_{L,R} =  \left(\ba{cc} W^U_{L,R} & 0 \\ 0 & W^D_{L,R} \ea\right) \,; \qquad
Q_{L,R} =  \left(\ba{cc} U_{L,R} & 0 \\ 0 & D_{L,R} \ea\right) \,.
\eea
Note that time-reversal invariance of the unrotated Hamiltonian $\CH'$
implies that $E(W,Q) = E(W^*,Q^*)$. Hence, spontaneous CP violation occurs if
the solutions $W_0$, $Q_0$ to the minimization problem are not real (up to an
overall $\CZ_N$ phase).

In Eq.~(\ref{eq:vacE}), $\Delta_{TT}$, $\Delta_{Tq}$ and $\Delta_{qq}$ are
{\em positive} four-fermion condensates in the standard vacuum, $\rvac$. They
are renormalized at the appropriate $\Metc$ scale and are given approximately
by
\bea\label{eq:largeNconds}
\Delta_{TT} &\simeq& (\Delta_T(\Metc))^2 \nn\\
\Delta_{Tq} &\simeq& \Delta_T(\Metc) \, \Delta_q(\Metc)\\
\Delta_{qq} &\simeq& (\Delta_q(\Metc))^2\,.\nn
\eea
In walking technicolor (see Appendix~A)
\be\label{eq:tccond}
\Delta_T(\Metc) \simle (\Metc/\Lambda_{TC}) \, \Delta_T(\Lambda_{TC}) =
10^2-10^4 \times \Delta_T(\Lambda_{TC})\,.
\ee
In QCD, however,\footnote{In TC2, Eq.~(\ref{eq:qcdcond}) must be modified to
  account for the embedding of $\suc$ into $\suone\otimes\sutwo$ and the
  latter group's embedding into $\Getc$.}
\be\label{eq:qcdcond}
\Delta_q(\Metc) \simeq (\log(\Metc/\Lambda_{QCD}))^{\gamma_m} \,
\Delta_q(\Lambda_{QCD}) \simeq \Delta_q(\Lambda_{QCD})\,,
\ee
where the anomalous dimension $\gamma_m$ of $\ol q q$ is small.\footnote{We
  shall assume that $\gamma_m$ remains small in FUTC2, even though quarks
  have strong $\suone$ interactions there.} Thus,
\be\label{eq:ratio}
\rho = \frac{\Delta_{Tq}(\Metc)}{\Delta_{TT}(\Metc)} \simeq
\frac{\Delta_{qq}(\Metc)}{\Delta_{Tq}(\Metc)} \simeq
\frac{\Lambda_{TC}}{\Metc} \left(\frac{f_\pi}{F_T}\right)^3 \simle 10^{-11} 
\ee
for $F_T \simeq 100\,\gev$. This ratio is $10^2$--$10^4$ times smaller than
it is in a technicolor theory in which the coupling does not walk.

The last line of Eq.~(\ref{eq:vacE}) makes clear that we should first
minimize the energy $E_{TT}$ in the technifermion sector. Because $W$ may be
assumed block-diagonal, $E_{TT}$ factorizes into two terms, $E_{UU} +
E_{DD}$, in which $W_U$ and $W_D$ may each be taken unimodular. We may then
minimize separately in the $\CU$ and $\CD$ sectors. This determines
$W_0=(W_0^U,W_0^D)$ , and as we shall see, $\ol\theta_q$, up to corrections
of $\CO(10^{-11})$ from the quark sector.\footnote{Two other sorts of
  corrections need to be studied. The first are higher-order ETC and
  electroweak contributions to $E_{TT}$. The electroweak ones are naively
  $\CO(10^{-7})$, much too large for $\ol\theta_q$. The second are due to
  $\ol T t \ol t T$ terms in $E_{Tq}$ which may be important if the top
  condensate is large. I thank J.~Donoghue and S.~L.~Glashow for emphasizing
  the potential importance of these corrections.} This result is then fed
into $E_{Tq}$ which is minimized to determine $Q_0$ --- and the nature of {\em
  weak} CP violation in the quark sector --- up to corrections which are also
$\CO(10^{-11})$.

In Ref.~\cite{Lane:2000es}, it was shown that minimizing $E_{TT}$ leads to
three possibilities for the phases in $W$. (We drop its subscript ``0'' from
now on.) Let us write $W_{IJ} = |W_{IJ}| \exp{(i\phi_{IJ})}$. Consider an
individual term, $-\Lambda^{TT}_{IJKL} \, W_{JK} \, W^\dag_{LI} \,
\Delta_{TT}$, in $E_{TT}$.  If $\Lambda^{TT}_{IJKL} > 0$, this term is least
if $\phi_{IL} = \phi_{JK}$; if $\Lambda^{TT}_{IJKL} < 0$, it is least if
$\phi_{IL} = \phi_{JK} \pm \pi$. We say that $\Lambda^{TT}_{IJKL} \neq 0$
{\em links} $\phi_{IL}$ and $\phi_{JK}$, and tends to align (or antialign)
them. Of course, the constraints of unitarity may partially or wholly
frustrate this alignment. The three cases for the phases $\phi_{IJ}$ are:

\begin{enumerate}

\item{} The phases are all unequal, irrational multiples of $\pi$ that are
random except for the constraints of unitarity and unimodularity.

\item{} All of the phases are equal to the same integer multiple of $2\pi/N$
  (mod~$\pi$). This may occur when all phases are linked and aligned, and the
  value $2\pi/N$ is a consequence of the unimodularity of $W_U$ and $W_D$. In
  this case we say that the phases are ``rational''.
  
\item{} Several groups of phases may be linked among themselves but not with
  others. The phases may then be only partially aligned and they take on {\em
    various} rational multiples of $\pi/N'$ for one or more integers $N'$
  from~1 to~$N$.

\end{enumerate}

\nin As far as we know, such nontrivial rational phases (i.e., $\neq
0,\pi,\pi/2$) occur naturally only in ETC theories. They are a consequence of
$E_{TT}$ being quadratic, not linear, in $W$ and of the instanton-induced
unimodularity constraints on $W$. Given these three possibilities, we now
investigate the quarks' CP violation.

There are two kinds of CP violation. Weak CP violation enters the standard
weak interactions through the CKM phase $\delta_{13}$ and the ETC and TC2
interactions through physically observable combinations of phases in the
quark alignment alignment matrices $Q_{L,R}$. Strong CP violation, which can
produce electric dipole moments $10^{10}$ times larger than the experimental
bound, is a consequence of instantons~\cite{Peccei:1996ax}. No theory of
the origin of CP violation is complete which does not eliminate strong CP
violation. Resolving this problem amounts to achieving $\ol\theta_q
\simle 10^{-10}$ {\em naturally}. Let us see how this might happen in
technicolor.

The ``primordial'' quark mass matrix element $(\CM_q)_{ij}$, the coefficient
of the bilinear $\ol q'_{Ri} q'_{Lj}$ of quark {\em electroweak} eigenstates,
is generated by ETC interactions and is given by\footnote{The matrix element
  $(\CM_u)_{tt}$ arises almost entirely from the TC2-induced condensation of top
  quarks. We assume that $\langle \ol t t \rangle$ and $(\CM_u)_{tt}$ are real in
  the basis in which $\theta_{QCD} = 0$. Since technicolor, color, and
  topcolor groups are embedded in ETC, all CP-conserving condensates are real
  in this basis.}
\be\label{eq:primordial}
 (\CM_q)_{ij} = \sum_{I,J} \Lambda^{Tq}_{IijJ} \, W^\dag_{JI} \,
 \Delta_T(\Metc)  \qquad (q,T = u,U \,\, {\rm or} \,\, d,D)\,.
\ee
The $\Lambda^{Tq}_{IijJ}$ are real ETC couplings of order
$(10^2$--$10^4\,\tev)^{-2}$ (see Appendix~A). Since the quark alignment
matrices $Q_{L,R}$ which diagonalize $\CM_q$ to $M_q$ are unimodular,
$\arg\det(M_q) = \arg\det(\CM_q) = \arg\det(\CM_u) + \arg\det(\CM_d)$.
Therefore, strong CP violation depends {\em entirely} on the character of
vacuum alignment in the technifermion sector --- the phases $\phi_{IJ}$ of
$W$ --- and by how the ETC factors $\Lambda^{Tq}_{IijJ}$ map these phases into
the $(\CM_q)_{ij}$.

If the $\phi_{IJ}$ are random irrational phases, $\ol \theta_q$ could vanish
only by the most contrived, unnatural adjustment of the $\Lambda^{Tq}$ and,
so, this case generically exhibits strong CP violation. If all $\phi_{IJ} =
2m\pi/N$ (mod $\pi$), then all elements of $\CM_u$ have the same phase, as do
all elements of $\CM_d$. Then, $U_{L,R}$ and $D_{L,R}$ will be real
orthogonal matrices, up to an overall phase. There may be strong CP
violation, but there will no weak CP violation in any interaction.

There remains the possibility, which we assume henceforth, that the
$\phi_{IJ}$ are different rational multiples of $\pi$. Then, strong CP
violation will be absent {\em if} the $\Lambda^{Tq}$ map these phases onto
the primordial mass matrix so that (1) each element $(\CM_q)_{ij}$ has a
rational phase {\em and} (2) these add to zero in $\arg\det(\CM_q)$. In the
absence of an explicit ETC model, we do not know whether this can happen, but
we see no reason it cannot. For example, there may be just one pair $(IJ)$
for which $\Lambda^{Tq}_{IijJ}\neq 0$ for fixed $(ij)$. An ETC model which
achieves such a phase mapping would solve the strong CP problem, i.e., $\ol
\theta_q \simle 10^{-11}$, without an axion and without a massless up quark.
This is, in effect, a ``natural fine-tuning'' of phases in the quark mass
matrix: rational phase solutions are stable against substantial changes in
the nonzero $\Lambda^{TT}$. There is, of course, no reason weak CP violation
will not occur in this scenario.

Determining the quark alignment matrices $Q_{L,R}$ begins with minimizing the
vacuum energy (using $\Delta_{Tq} \cong \Delta_T\Delta_q$)
\be\label{eq:ETq}
E_{Tq}(Q) \cong - {\rm Tr}\left(\CM_q \, Q + {\rm h.c.}\right)
\Delta_q(\Metc)
\ee
to find $Q=Q_L Q^\dag_R$. When $\ol\theta_q =0$, this is equivalent to making
the mass matrix diagonal, real, and positive. Whether or not $\ol \theta_q =
0$, the matrix $\CM_q Q$ is hermitian up to the identity
matrix~\cite{Dashen:1971et},
\be\label{eq:nuyts}
 \CM_q \, Q - Q^\dag \CM^\dag_q = i\nu_q \, 1 \,,
\ee
where $\nu_q$ is the Lagrange multiplier associated with the unimodularity
constraint on $Q$~\cite{Nuyts:1971az}, and $\nu_q$ vanishes if $\ol \theta_q$
does. Therefore, $\CM_q \, Q$ may be diagonalized by the single unitary
transformation $Q_R$ and, so,\footnote{Since quark vacuum alignment is based
  on first order chiral perturbation theory, it is inapplicable to the heavy
  quarks $c,b,t$.  However, since it just amounts to diagonalizing $\CM_q$
  when $\ol\theta_q = 0$, it correctly determines the quark unitary matrices
  $U_{L,R}$ and $D_{L,R}$ and the magnitude of strong and weak CP violation.}
\be\label{eq:Mdiag}
M_q \equiv \left(\ba{cc} M_u & 0 \\ 0 & M_d \ea\right) = Q^\dag_R \CM_q
\, Q Q_R = Q^\dag_R \CM_q \, Q_L \,.
\ee

The CKM matrix is $V= U^\dag_L D_L$. Carrying out the {\em vectorial} phase
changes on $q_{L,R \, i}$ required to put $V$ in the standard form with a
single CP-violating phase $\delta_{13}$, one
obtains~\cite{Harari:1986xf,Eidelman:2004wy}
\bea\label{eq:CKMmat}
V  &\equiv& \left(\ba{ccc}
      V_{ud} & V_{us} & V_{ub}\\
      V_{cd} & V_{cs} & V_{cb}\\
      V_{td} & V_{ts} & V_{tb}\\
      \ea\right)\\ \nn\\
 &=&  \left(\ba{ccc}
      c_{12\,} c_{13} 
      & s_{12\,} c_{13}
      & s_{13\,}e^{-i\delta_{13}}\\ 
      -s_{12\,}c_{23}-c_{12\,}s_{23\,}s_{13\,} e^{i\delta_{13}}
      & c_{12\,}c_{23}-s_{12\,}s_{23\,}s_{13\,} e^{i\delta_{13}}
      & s_{23\,}c_{13\,} \\
      s_{12\,}s_{23}-c_{12\,}c_{23\,}s_{13\,} e^{i\delta_{13}}
      & -c_{12\,}s_{23}-s_{12\,}c_{23\,}s_{13\,} e^{i\delta_{13}}
      & c_{23\,}c_{13\,}\\
      \ea\right) \,.\nn
\eea
Here, $s_{ij} = \sin\theta_{ij}$, and the angles $\theta_{12}$,
$\theta_{23}$, $\theta_{13}$ lie in the first quadrant. Additional
CP-violating phases appear in $U_{L,R}$ and $D_{L,R}$; certain linear
combinations are observable in the ETC and TC2 interactions. We discuss next
the form of $\CM_u$, $\CM_d$ and $U_{L,R}$, $D_{L,R}$ imposed by experimental
constraints on ETC and TC2.

First, limits on FCNC interactions, especially those contributing to $\Delta
M_K = M_{K_L}$--$M_{K_S}$ and the CP-violation parameter $\epsilon$, require
that ETC bosons coupling to the two light generations have masses
$\Metc/\getc \simge 10^3\,\tev$ (see Ref.~\cite{Lane:2002wv} for the latest
estimates). These can produce quark masses less than about $m_s(\Metc) \simeq
100\,\mev$ in a walking technicolor theory (see Fig.~\ref{fig:CPVB_fig_14} in
Appendix~A). We expect similar or smaller masses in the first two rows of
$\CM_{u,d}$ (except $\CM_{cc} \simeq 1\,\gev$). Extended technicolor bosons
as light as 50--$100\,\tev$ are needed to generate $m_b(\Metc) \simeq
4\,\gev$. Flavor-changing neutral current interactions mediated by such light
ETC bosons must be suppressed by small mixing angles between the third and
the first two generations.\footnote{We must assume that the ETC interactions
  $\CH'_{qq}$ are electroweak generation conserving to suppress $|\Delta S|
  =2$ FCNC interactions adequately. We also assume that the magnitude of
  $\Lambda^{qq}_{ijkl}$ is comparable to that of $\Lambda^{Tq}_{iILl}$ and/or
  $\Lambda^{Tq}_{JjkK}$. See Eq.~(\ref{eq:gencon}) in Sect.~3.}

The most important feature of $\CM_u$ is that the TC2 component of
$(\CM_u)_{tt}$, $\hat m_t = 160$--$170\,\gev$, is much larger than its other
elements, all of which are generated by ETC exchange. In particular,
off-diagonal elements in the third row and column of $\CM_u$ are expected to
be no larger than the 0.01--1.0~GeV associated with $m_u$ and $m_c$. So,
$\CM_u$ and $U_{L,R}$ are very nearly block-diagonal with $|U_{L,R \, t
  u_i}| \cong |U_{L,R \, u_i t}| \cong \delta_{t u_i}$.

There is an argument that the matrix $\CM_d$ must have, or nearly have, a
triangular texture~\cite{Lane:1996ua}; also see Ref.~\cite{Buchalla:1996dp}:
In 1995 Kominis argued that in topcolor models the $\suone\otimes \uone$
couplings of the bottom quark are not far from the critical values required
for condensation. Consequently, there ought to exist so-called ``bottom
pions'' --- relatively light ($\simeq 300\,\gev$) scalar bound states of $\ol
t_L b_R$ and $\ol b_L b_R$ that couple strongly ($\propto \hat m_t$) to third
generation quarks~\cite{Kominis:1995fj}. Bottom pions will induce excessive
$B_d $--$\ol B_d$ mixing unless $|D_{Lbd} D_{Rbd}| \simle 10^{-7}$. In
addition to this, since $U_L$ is block-diagonal, the observed CKM mixing
between the first two generations and the third must come from the down
sector matrix $D_L$. These considerations (and the need for flavor symmetry
in the two light generations) imply that the $\CM_d$ is approximately
triangular, with its $d_R,s_R \leftrightarrow b_L$ elements $\CM_{db}$ and
$\CM_{sb}$ much smaller than its $d_L,s_L \leftrightarrow b_R$ elements.

A triangular $\CM_d$ was produced in the TC2 models of
Refs.~\cite{Lane:1995gw, Lane:1996ua,Lane:1998qi} by choosing the topcolor
$\uone$ charges to forbid ETC interactions that induce the $d_R,s_R
\leftrightarrow b_L$ matrix elements $\CM_{db}$, $\CM_{sb}$. Then $D_R$, like
$U_{L,R}$, is nearly $2\times 2$ times $1\times 1$ block-diagonal. Although
both $V_8$ and $Z'$ exchange produce FCNC interactions, this constraint plus
the approximate flavor symmetry implied by Eq.~(\ref{eq:YGIM}) mean that many
interesting FCNC effects arise only from left-handed $\ol b'_L b'_L \ol b'_L
b'_L$ TC2 interactions. Then the magnitudes and phases of the mixing factors
are simply related to those of CKM elements (as in Eq.~(\ref{eq:ckm}) below).

Kominis' argument for near-criticality of $b$-quark TC2 interactions relied
on assuming it had standard-model $\uone$ hypercharges $Y_{1Lb} = 1/6$ and
$Y_{1Rb} = -1/3$ {\em and} that the $\uone$ coupling is not very strong (to
avoid a Landau pole at low energy). As we noted above, however, a strong
$\uone$ coupling is needed to avoid fine-tuning the $\suone$ coupling, while
the Landau pole might be avoided if $\uone$ is embedded in $\Getc$ at low
enough energy. Thus, we view the existence of bottom pions as arguable at
best. For STC2, we shall consider both cases: $D_R$ is block-diagonal and it
is not.

In Ref.~\cite{Simmons:2001va} Simmons pointed out that there may be an
amelioration of the bottom pion contribution to $B_d$--$\ol B_d$ mixing
in FUTC2. There, all quarks have strong attractive $\suone$ interactions so
that there would be ``q-pions'' with flavor-symmetric couplings for all the
quarks {\em if} $\uone$ is relatively weak --- as Simmons assumed necessary
to avoid a Landau pole.  Such q-pions would not induce $B_d$--$\ol B_d$
mixing. However, whether or not light quarks transform under $\uone$, the
heavy ones do {\em and} their $\uone$ couplings should not be weak. This
ruins the flavor symmetry of q-pions' couplings and Simmons' argument fails.
Furthermore, if light quarks do have $\uone$ interactions then, like the
bottom quark's, they probably must be repulsive. This calls into question the
very existence of {\em all} q-pions {\em except} the top pion.  In short, the
matter of q-pions and their induced $B_d$--$\ol B_d$ mixing is highly
uncertain. In FUTC2, even more than in STC2, the argument for $\CM_d$ to be
triangular and $D_R$ block-diagonal is weak and, like Simmons, we shall not
be bound by it.

Because mixing between the third generation and the two lighter ones comes
from $D_L$, in either TC2 variant we have (using $V_{tb} \cong 1$, and
independent of the form of $D_R$)
\be\label{eq:ckm}
 V_{td_i} \cong V^*_{tb} \, V_{td_i} \cong U_{L tt} D^*_{L bb}
  U^*_{L tt}  D_{L bd_i} \cong D^*_{L bb} D_{L bd_i} \,.
\ee
This relation is good to 5\%. Together with the assumption that $D_R$ is
block-diagonal, it was used in Ref.~\cite{Burdman:2000in} to put limits on
the TC2 $V_8$ and $Z'$ masses from $B_d$--$\ol B_d$ mixing. It was shown
there that $M_{V_8}$, $M_{Z'} \simge 5\,\tev \gg \hat m_t$. This implies that
the TC2 gauge couplings must be tuned to within 1\% or better of their
critical values --- an uncomfortably fine tuning in a dynamical theory. For
FUTC2, Simmons' q-pion argument and assumption that $\uone$ is relatively
weak led her to conclude that this bound could be lowered for the $Z'$ (as
well as for the $V_8$ which does not mediate FCNC interactions). We revisit
this question in Sect.~3 and conclude that the bound $M_{V_8,\,Z'} \simge
5\,\tev$ generally holds in both TC2 variants.

To calculate the TC2 and ETC contributions to CP-violating parameters in $K$
and $B$-decays, we generated three representative sets of alignment matrices
$U_{L,R}$ and $D_{L,R}$. They were created by carrying out vacuum alignment
with a non-hermitian primordial mass matrix $\mathcal M_q$ satisfying
$\arg\det(\CM_q) = 0$.  The first set of alignment matrices (Mass Model~1)
has a block-diagonal $D_R$, as was assumed
Ref.~\cite{Burdman:2000in}.\footnote{Sets of alignment matrices were created
  in this way in Refs.~\cite{Lane:2001rv,Lane:2002wv} to calculate the TC2
  and ETC contributions to $\epsilon$.} Mass Models~2 and~3 have $D_R \sim
D_L$. The mass and alignment matrices are presented in Appendix~B. As we
shall see in Sect.~5, a discrepancy in the value of $\sin{2\beta}$ measured
in different decays is possible only in models with a non-block-diagonal $D_R$.

\section*{3. TC2 and ETC Interactions}

At energies well below $M_{V_8,\,Z'}$, the effective TC2 current $\times$ current
interaction is
\be\label{eq:Htctwo}
 \CH_{TC2} \equiv  \CH_{Z'} + \CH_{V_8} 
  = \frac{g^2_Y}{2M^2_{Z'}}J_{Z'\, \mu}J_{Z'}^{\mu} +
    \frac{g^2_C}{2M^2_{V_8}}J^{A\,\mu}J^A_{\mu}\,,
\ee
where
\bea\label{eq:currents}
       J_{Z'\,\mu} & = & \sum_{\lambda = L,R}\sum_{i}\Big
       (Y_{1\lambda i}\cot{\theta_Y} - Y_{2\lambda i}\tan{\theta_Y} \Big )\,
       \bar q_{\lambda i}'\gamma_\mu q_{\lambda i}'\,,\nn\\
       J^A_\mu & = & \sum_{\lambda = L,R}\Big(\sum_{i = t,
       b}A_h - \sum_{i = u,d,s,c}A_l \Big)\,
     \bar q_{\lambda i}'\gamma_\mu \frac{\lambda_A}{2} q_{\lambda i}' \,.
\eea
The primed fields are electroweak eigenstates. The couplings $g_Y$ and $g_C$
are the standard model hypercharge and color couplings, defined in terms of
the original $SU(3)$ ($U(1)$) couplings by
\bea\label{eq:couplings}
g_C = \frac{g_1g_2}{\sqrt{g^2_1 + g^2_2}} &;& \frac{g_1}{g_2} =
\cot{\theta_C}  \gg 1 \nonumber \\
g_Y = \frac{g'_1g'_2}{\sqrt{g'^2_1 + g'^2_2}} &;& \frac{g'_1}{g'_2} =
\cot{\theta_Y} \gg 1 \nonumber 
\eea
The $\uone$ and $\utwo$ hypercharges $Y_{1\lambda i}$ and $Y_{2\lambda i}$
satisfy the flavor symmetry conditions in Eq.~({\ref{eq:YGIM}) and, of
  course, $Y_{Li} = Y_{1Li} + Y_{2Li} = 1/6$ and $Y_{Ri} = Y_{1Ri} + Y_{2Ri}
  = e_i$, the electric charge of $q_i$. We shall ignore the $Y_{2\lambda
    i}\tan{\theta_Y}$ terms in our calculations.  The couplings of the heavy
  and light quarks to the coloron, $A_h$ and $A_l$, depend on the TC2 model.
  In STC2, only the third generation couples to the strong $\suone$, so that
  $A_h = \cot{\theta_C}$ and $A_l = \tan{\theta_C}$.  In this case, both
  $\CH_{Z'}$ and $\CH_{V_8}$ contain FCNC interactions.  In FUTC2, all quarks
  have the same coupling to the colorons, $A_h = -A_l = \cot{\theta_C}$, and
  only $\CH_{Z'}$ has FCNC interactions.
   
  Expanding $\CH_{TC2}$ for the FUTC2 and STC2 variants, and keeping only the
  strongly coupled $U(1)_1$ and $SU(3)_1$ contributions to potential FCNC
  interactions, we obtain
\bea\label{eq:FUandS}
       \CH_{FU} &=& \CH_{Z'} =
         \frac{g^2_Y\cot^2{\theta_Y}}{2M^2_{Z'}}\sum_{i,j}\sum_{\lambda_1,
         \lambda_2 = L,R}Y_{\lambda_1 i}Y_{\lambda_2 j}\,
         \bar{q}_{\lambda_1 i}'\gm q_{\lambda_1 i}'\,
         \bar{q}_{\lambda_2 j}'\gamma_{\mu} q_{\lambda_2 j}' \,;\nn\\ \\
\CH_{S} &=& \CH_{Z'} + \CH_{V_8} = \CH_{FU} +
  \frac{g^2_C\cot^2{\theta_C}}{2M^2_{V_8}}\sum_{i,j = t,b}
  \sum_{\lambda_1, \lambda_2 =L,R}
  \bar q_{\lambda_1 i}'\gm \frac{\lambda_A}{2} q_{\lambda_1 i}'\,
  \bar q_{\lambda_2 j}'\gamma_{\mu}\frac{\lambda_A}{2} q_{\lambda_2 j}' \,.\nn
\eea
>From now on, we denote $Y_{1\lambda i}$ by $Y_{\lambda i}$. Remember that we
assume that all quarks transform nontrivially under the strong $U(1)_1$.

Still assuming that the ETC gauge group commutes with electroweak $SU(2)$,
the ETC four-quark interaction to lowest order in $G^2_{ETC}$ is
\bea\label{eq:HETC}
\CH_{ETC} \equiv \CH'_{qq} &=& \Lambda^{LL}_{ijkl} \left(\ol u'_{Li} \gamma^\mu
  u'_{Lj} + \ol d^{\ts \prime}_{Li} \gamma^\mu d^{\ts \prime}_{Lj}\right)
\left(\ol u'_{Lk} \gamma^\mu u'_{Ll} + \ol d^{\ts \prime}_{Lk} \gamma^\mu
  d^{\ts
    \prime}_{Ll}\right) \nn\\
& & \ts +
\left(\ol u'_{Li} \gamma^\mu u'_{Lj} + \ol d^{\ts \prime}_{Li} \gamma^\mu 
d^{\ts \prime}_{Lj}\right)
\left(\Lambda^{u,LR}_{ijkl} \ol u'_{Rk}\gamma^\mu u'_{Rl} +
      \Lambda^{d,LR}_{ijkl} \ol d^{\ts \prime}_{Rk}\gamma^\mu
      d^{\ts \prime}_{Rl}\right) \nn \\
& & \ts + \Lambda^{uu,RR}_{ijkl}\ol u'_{Ri}\gamma^\mu u'_{Rj}\ts\ol
u'_{Rk}\gamma^\mu u'_{Rl} + \Lambda^{dd,RR}_{ijkl}\ol d^{\ts
\prime}_{Ri}\gamma^\mu d^{\ts \prime}_{Rj} \ts\ol d^{\ts
\prime}_{Rk}\gamma^\mu d^{\ts \prime}_{Rl} \nn \\
& & \ts + \Lambda^{ud,RR}_{ijkl} \ol u'_{Ri}\gamma^\mu u'_{Rj}\ts
\ol d^{\ts \prime}_{Rk}\gamma^\mu d^{\ts \prime}_{Rl} \ts.
\eea
Since the ETC gauge group contains technicolor, color and topcolor, and
flavor as commuting subgroups, the flavor currents in $\CH_{ETC}$ are color
and topcolor singlets. The operators are renormalized at the ETC scale of
their $\Lambda$-coefficients. Hermiticity and CP-invariance of $\CH_{ETC}$
implies that $\Lambda_{ijkl} = \Lambda^*_{jilk} = \Lambda^*_{ijkl}$.  When
written in terms of mass eigenstate fields $q_{L,R\ts i} = \sum_j
(Q^\dag_{L,R})_{ij} q'_{L,R\ts j}$ with $Q = U,D$, an individual term in
$\CH_{ETC}$ has the form
\be\label{eq:Hqterm}
  \left(\sum_{i'j'k'l'} \Lambda^{q_1 q_2 \lambda_1 \lambda_2}_{i'j'k'l'}
  \ts\ts   Q^\dag_{\lambda_1\ts ii'} Q_{\lambda_1\ts j'j} \ts
  Q^\dag_{\lambda_2\ts kk'} Q_{\lambda_2\ts l'l}\right) \ts 
  \ol q_{\lambda_1 i} \ts \gamma^\mu \ts q_{\lambda_1 j} \ts\ts
  \ol q_{\lambda_2 k} \ts \gamma_\mu \ts q_{\lambda_2 l} \ts. 
\ee

The $\Lambda$'s in $\CH_{ETC}$ are of order $\getc^2/\Metc^2$. A reasonable
(and time-honored) guess for the magnitude of the $\Lambda_{ijkl}$ is that
they are comparable to the ETC coefficients that generate the quark mass matrix
$\CM_q$. To estimate the FCNC in $\CH_{ETC}$, we elevate this to a rule: The
ETC scale $\Metc/\getc$ in a term involving weak eigenstates of the form $\ol
q^{\ts \prime}_i q'_j \ol q^{\ts \prime}_j q'_i$ or $\ol q^{\ts \prime}_i
q'_i \ol q^{\ts \prime}_j q'_j$ (for $q'_i = u'_i$ or $d^{\ts \prime}_i$) is
approximately the same as the scale that generates the $\ol q^{\ts
  \prime}_{Ri} q'_{Lj}$ mass term, $(\CM_q)_{ij}$. A plausible, but
approximate, scheme for correlating a quark mass $m_q(\Metc)$ with
$\Metc/\getc$ is presented in Appendix~A (see Fig.~\ref{fig:CPVB_fig_14}).

Extended technicolor masses, $\Metc/\getc \simge 1000\,\tev$, are necessary,
but not sufficient, to suppress FCNC interactions of light quarks to an
acceptable level. Without further suppression by CKM-like mixing angles, the
ETC masses required for compatibility with $\epsilon$ are so large that, even
with walking technicolor, light-quark masses are too
small~\cite{Lane:2002wv}. Thus, we need to assume that $\CH_{ETC}$ is {\it
  electroweak generation conserving}, i.e.,
\be\label{eq:gencon}
\Lambda^{q_1q_2\lambda_1\lambda_2}_{ijkl} = \delta_{il}\delta_{jk}
\Lambda^{q_1q_2\lambda_1\lambda_2}_{ij} +  \delta_{ij}\delta_{kl}
\Lambda^{\prime \ts q_1q_2\lambda_1\lambda_2}_{ik}\ts.
\ee
Considerable FCNC suppression then comes from off-diagonal elements in the
alignment matrices $Q_{L,R}$.

Note that the TC2 and ETC interactions generally have RL $((V + A)\times(V -
A))$ and RR $((V + A)\times(V + A))$ ``wrong chirality'' structure as well as
the LL $((V - A)\times (V - A))$ and LR $((V - A)\times(V + A))$ structure
found in standard model contributions to FCNC interactions.

\section*{4. Constraints on the TC2 and ETC Interactions} 

The first constraint we consider is that which top quark condensation, but
not not bottom nor light quark condensation, places on the TC2 couplings and
hypercharges. In the Nambu--Jona-Lasinio approximation, a $\ol qq$ condensate
occurs when the quark's couplings satisfy
\be\label{eq:critical}
 \alpha_q(V_8) + \alpha_q(Z') \equiv \frac{g^2_C A^2_q}{3\pi^2} +
\frac{g^2_Y\cot^2{\theta_Y}(Y_{Lq}Y_{Rq})}{4\pi^2} \ge 1.
\ee
The so-called critical values of the couplings occur when the equality is
satisfied. As we have stressed, both terms should be $\CO(1)$ to avoid fine
tuning. In STC2, this strongly suggests that
\be\label{eq:stccon}
Y_{Lt}Y_{Rt} > 0\,; \quad Y_{Lb}Y_{Rb} < 0  \qquad{\rm (STC2)}\,.
\ee
Because $A_l^2 = \tan^2\thc \ll 1$, the constraint on light quarks is rather
loose, however:
\be\label{lightcon}
g^2_Y\cot^2\thy \,Y_{Lq}Y_{Rq}/4\pi^2 < 1\qquad{\rm (STC2)}\,.
\ee
In FUTC2, the condition that only the top quark condenses is most simply met
by requiring that, for all quarks except top,
\be\label{futccon}
Y_{Lq}Y_{Rq} < 0\qquad\qquad\qquad\qquad{\rm (FUTC2)}\,.
\ee
We assume this from now on.
  
Other limits on the couplings and masses in $\CH_{TC2}$ and $\CH_{ETC}$ come
from mixing and CP violation in the $K^0$ and $B_d$ meson systems. The
constraint from the kaon $\epsilon$ parameter for models in which $D_R$ is
block diagonal were discussed in Refs.~\cite{Lane:2001rv,Lane:2002wv}. For
$\Lambda_{ssss} \simeq (2000\,\tev)^{-2}$ and $M_{V_8} \simeq M_{Z'} \simeq
10\,\tev$, it was shown there it is not difficult to account for the measured
value of $\epsilon$. For models with a nontrivial $D_R$ and these mass
scales, the $\epsilon$ parameter is not a strong constraint at all. Varying
the relative strengths and signs of $\Lambda_{ssss}$ and $\Lambda'_{ssss}$
can cause changes of up to $\pm 100$ times the standard model $\epsilon$. The
more incisive constraint on TC2 --- but not on ETC --- comes from
$\BDbar$--$\BD$ mixing.  This was considered first (mainly for STC2) in
Ref.~\cite{Burdman:2000in} and reconsidered (especially for FUTC2) in
Ref.~\cite{Simmons:2001va}. We reconsider both TC2 variants in this section.

The $B_H^0$--$B_L^0$ mass difference $\Delta M_{B_d} = (3.22 \pm 0.05)\times
10^{-10}\,\mev $ is directly related to the off-diagonal matrix element
$M_{12}$ of the $\BDbar - \BD$ Hamiltonian~\cite{Buras:2001pn}. Since
$|\Gamma_{12}| \ll |M_{12}|$, we have $\Delta M_{B_d} = 2|M_{12}|$. The
standard model contributions to $M_{12}$ come from box diagrams which are
proportional to $V_{td}^2$ and therefore carry a CKM phase $-2\beta$. New
physics contributions, at tree and loop levels, can alter the magnitude and
phase of $M_{12}$. However, $M_{12}$-mixing occurs in all neutral $B$ decays,
so that new physics in mixing alone {\em cannot} explain the $\sin{2\beta}$
discrepancy (see Sect.~6).

In both FUTC2 and STC2, the dominant new contribution to $M_{12}$ comes from
$\ol b'b' \ol b'b'$ terms in $\CH_{TC2}$. For FUTC2, the interaction is
\begin{eqnarray}\label{eq:HMFU}
  \CH_{FU}(M_{12}) &=&\frac{g_Y^2\cot^2{\theta_Y}}{8M^2_{Z'}}\Big[ (\Delta
  Y_L)^2(D_{Lbb}D^*_{Lbd})^2 (\bar d b)_{V - A}(\bar d b)_{V - A} \nn\\
  &&\qquad\qquad\,\, + (\Delta Y_R)^2(D_{Rbb}D^*_{Rbd})^2(\bar d b)_{V +
  A}(\bar d b)_{V + A}\\
  &&\qquad\qquad\,\, +2(\Delta Y_L)(\Delta Y_R)(D_{Lbb}D^*_{Lbd}D_{Rbb}D^*_{Rbd})
  (\bar d b)_{V - A}(\bar d b)_{V + A} + {\rm h.c.}\,\Big],\nn
\end{eqnarray}
where $(\bar d b)_{V \pm A} = \ol d\gamma_\mu(1\pm \gamma_5)b$ and the
appearance of $\Delta Y_{\lambda} = Y_{\lambda b} - Y_{\lambda d}$ reflects
the approximate flavor symmetry of Eq.~(\ref{eq:YGIM}). Then $M_{12}$ is
estimated in the vacuum insertion approximation to
be~\cite{Burdman:2000in,Simmons:2001va}:
 \begin{eqnarray}\label{eq:DelMFU}
 2(M_{12})_{FU} &=&
 \frac{g^2_Y\cot^2\thy}{3 M^2_{Z'}}F^2_{B_d}\hat B_{B_d} M^2_{B_d}
        \eta_B \Big{[}\Delta Y_L^2(D_{Lbb}D^*_{Lbd})^2 
 + \Delta Y_R^2 (D_{Rbb}D^*_{Rbd})^2 \nn\\
 &&\,\,\, - \Big(\frac{3}{2} +
        \frac{M_B^2}{(m_b + m_d)^2}\Big)
        \Delta Y_L\Delta Y_R(D_{Lbb}D^*_{Lbd}D_{Rbb}D^*_{Rbd})\Big{]} \,.
 \end{eqnarray}
For STC2 models, there is also a coloron contribution:
\begin{eqnarray}\label{eq:DelMS}
2(M_{12})_{S} &=& 2(M_{12})_{FU} + \frac{g^2_C\cot^2{\theta_C}}{9 M^2_{V_8}}
       F^2_{B_d}\hat B_{B_d} M^2_{B_d} \eta_B\\
&&\times \Big{[}(D_{Lbb}D^*_{Lbd})^2 + ({D_{Rbb}D^*_{Rbd})^2 
-\Big (\frac{3}{2} + \frac{M_B^2}{(m_b + m_d)^2}\Big)
(D_{Lbb}D^*_{Lbd} D_{Rbb}D^*_{Rbd}}\Big{]} \,.\nn
\end{eqnarray}
Here, $\eta_B = 0.55 \pm 0.01$ is a QCD radiative correction factor for the
LL and RR product of color-singlet currents and we assume it to be the same
for the LR product. We take $F_{B_d} \sqrt{B_{B_d}} = (200\pm
40)\mev$~\cite{Buras:1998fb}, where $F_{B_d}$ and $B_{B_d}$ are,
respectively, the $B_d$--meson decay constant and bag parameter. The
additional factor of 1/3 in the coloron contribution comes from the Fierz
rearrangement to a product of color singlet currents. These TC2 contributions
to $M_{12}$ are added to the standard-model one~\cite{Buras:2001pn},
\be\label{eq:DelMSM}
2 (M_{12})_{SM} = \frac{G^2_F}{6\pi^2} \eta_B M_{B_d} F^2_{B_d}
M^2_W S_0(x_t) (V^*_{tb} V_{td})^2 \ts,
\ee
where the top-quark loop function $S_0(x_t) \cong 2.3$ for $x_t =
m_t^2(m_t)/M^2_W$ and $m_t(m_t) = 167\,\gev$. The TC2 contributions to
$M_{12}$ come from operators renormalized at $M_{Z'}$ and $M_{V_8}$ rather
than at $M_W$. For simplicity, we take $M_{Z'} = M_{V_8}$ unless stated
otherwise. We assume that operator renormalizations from $M_{Z'}$ to $M_W$
are simply multiplicative, $\CO(1)$, and can therefore be
ignored\footnote{Since TC2 contributions occur at tree level, $\Gamma_{12}$
  is unaffected. Therefore, we still have $|\Gamma_{12}| \ll |M_{12}|$, and
  the ratio $q/p$, which describes the degree of mixing in the physical
  eigenstates (in Eq.~(\ref{eq:lambdaCP}) below), is still a pure
  phase~\cite{Buras:2001pn}.}.
     
As Simmons pointed out, including the RR and LR operators in $M_{12}$ opens
the possibility of obtaining lower mass limits than in
Ref.~\cite{Burdman:2000in}. She found $M_{Z'} \simge 1\,\tev$ in FUTC2.
However, Simmons assumed smaller $\uone$ couplings than we do. Furthermore,
$D_R$ and $D_L$ matrix elements that lead to lower TC2 boson masses may not
arise from alignment with plausible mass models. This, in fact, is what we
found for the mass matrices considered in Appendix~B.

To set the mass limits, we followed Ref.~\cite{Burdman:2000in} in assuming
that the $\suone$ and $\uone$ couplings of the top quark are each half their
critical value, $\alpha_t(V_8) = \alpha_t(Z') = 1/2$, i.e., 
\begin{equation}\label{eq:halfcrit}
  g^2_C\cot^2{\theta_C} = \frac{3\pi^2}{2}\,,\quad
  g_Y^2\cot^2{\theta_Y} Y_{Lt}Y_{Rt} = 2\pi^2 \,.
\end{equation}
We also assumed $(\Delta Y_L)^2 \equiv (Y_{Lb} - Y_{Ld})^2 = Y_{Lt}Y_{Rt}$.
These assumptions are reasonable, given the need to avoid fine-tuning, but
the mass limits are somewhat sensitive to them. As noted above, we also
assumed $M_{V_8} = M_{Z'}$ for STC2. Because all the mixing matrix factors
are determined by the primordial quark mass matrix $\CM_q$, the only
remaining free parameters are the ratio of hypercharge differences, $\xi =
\Delta Y_R/\Delta Y_L$, and the gauge boson masses. Equating twice the total
$|M_{12}|$ to the measured $\Delta M_{B_d}$, we determined the gauge boson
mass limit as a function of $\xi$.  The lowest possible gauge boson masses
for the interval of $-5 < \xi < 5$ for STC2 and FUTC2 and the three sets of
alignment matrices are:
\begin{equation}\label{eq:MVZlimits}
\begin{array}{cccc} \underline{{\rm MASS}\ {\rm MODEL}} & \underline{{\rm
      TC2}\ {\rm MODEL}} &
  \underline{M_{min}} & \underline{{\rm FINE\ TUNING}} \\
         1 & {\rm STC2} &  23.9\,\tev & 0.05\% \\
         2 & {\rm STC2} &  \,\,\,5.0\,\tev & 0.9\% \\
         3 & {\rm STC2} &  \,\,\,10.5\,\tev & 0.2\% \\
           &      &  \\
         1 & {\rm FUTC2} &  21.3\,\tev & 3\% \\
         2 & {\rm FUTC2} &  \,\,\,7.0\,\tev & 3\% \\
         3 & {\rm FUTC2} &  \,\,\,10.1\,\tev & 3\% \\
       \end{array}
\end{equation}
The last column is an estimate of the fine tuning of the TC2 couplings; this is
discussed below.
 
In Mass Models~2 and~3, which produce $|D_{Rbq}| \simeq |D_{Lbq}|$, the
bounds on $M_{V_8,Z'}$ are lower than in Model~1, as Simmons anticipated.
However, this effect is not limited to FUTC2. Nor are the bounds as low as
Simmons determined because she assumed a relatively weak $\uone$ coupling and
$Y_{Lb} Y_{Rb} = -1/18$. Although Mass Model~1 satisfies the relationship
Eq.~(\ref{eq:ckm}) used in Ref.~\cite{Burdman:2000in}, the mass limits do not
agree. The disagreement is caused by using different values of $V_{td}$. The
mass limits scale approximately as $|V_{td}|$ in Model~1. In
Ref.~\cite{Burdman:2000in}, the minimal values $|V_{td}| = 0.005$ and 0.0034
were used. Here, we derived $V_{td}$ from $\CM_q$, obtaining $|V_{td}| =
.0075$ in Model~1 and $|V_{td}| = 0.0055$ in Models~2 and~3. As noted in
Ref.~\cite{Simmons:2001va}, once $D_R$ is no longer block-diagonal, the
$\Delta M_{B_d}$ constraint is no longer model-independent, i.e., determined
solely by the CKM element $V_{td}$. This is clearly demonstrated in the
factor of 1.5--2 difference in the bounds for Models~2 and~3.

Finally, producing a TC2 contribution $\hat m_t \simeq 165\,\gev$ to the top
quark mass with such large $Z'$ and $V_8$ masses implies fine tuning of the
couplings to their critical value~\cite{Chivukula:1996cc}. The fine tuning is
characterized by the magnitude of
\begin{equation}\label{eq:tuning}
  \frac 
  { \alpha_t(Z')\frac{\hat m^2_t}{M^2_{Z'}}\log{\Big(\frac{M^2_{Z'}}{\hat
  m^2_t}\Big)} + 
    \alpha_t(V_8)\frac{\hat m^2_t}{M^2_{V_8}}\log{\Big(\frac{M^2_{V_8}}{\hat
  m^2_t}\Big)}} {\alpha_t(Z')\Big[1-\frac{\hat
  m^2_t}{M^2_{Z'}}\log{\Big(\frac{M^2_{Z'}}{\hat m^2_t}\Big)}\Big] +
   \alpha_t(V_8)\Big[1-\frac{\hat
  m^2_t}{M^2_{V_8}}\log{\Big(\frac{M^2_{V_8}}{\hat m^2_t}\Big)}\Big]}
\end{equation}
In Ref.~\cite{Burdman:2000in}, with $M_{V_8} \simeq 5\,\tev$ and $M_{Z'}
\simeq 10\,\tev$, the fine tuning was found to be 0.5\% using the NJL
approximation with half-critical couplings. Using $M_{Z'} = M_{V_8}$ with the
appropriate STC2 mass limits, we obtain the fine tuning estimates $\simle
1\%$ listed in Eq.~(\ref{eq:MVZlimits}). In FUTC2, the situation is somewhat
better {\em if} we lower $M_{V_8}$ to the limit allowed by precision
electroweak observables, $M_{V_8}\simeq 1.6\ \TeV$~\cite{Popovic:1998vb}.
This leads to fine tuning of $3\%$. Fine tuning is also ameliorated if we
allow $(\Delta Y_L)^2 < Y_{Lt}Y_{Rt}$. This allows a lower $M_{Z'}$.
Obviously, this difference in the hypercharges cannot be too extreme. In
summary, despite more general assumptions on the structure of the alignment
matrices, we find the couplings to be as fine-tuned as in
Ref.~\cite{Burdman:2000in}. The principal reason is that we insist on a large
$\uone$ coupling to avoid fine tuning $\alpha_t(V_8)$!

\section*{5. $\BD \ra J/\psi K_S$, $\phi K_S$, $\eta' K_S$ and $\pi K_S$}

With the assumption that ETC interactions are generation conserving, their
contributions to $\BD$ decays are suppressed by small mixing angles and hence
negligibly small. The TC2 contributions to $b\ra s\bar q q$ decays are
obtained by writing Eqs.~(\ref{eq:FUandS}) in terms of mass eigenstates and
making use of the unitarity of the $D_{L,R}$:
\bea\label{eq:FUandSbsqq}
   \CH_{FU} &=&
   \frac{g^2_Y\cot^2{\theta_Y}}{2M^2_{Z'}}\sum_{\lambda_1,\lambda_2
   = L,R}D^*_{\lambda_1 bs}D_{\lambda_1 bb}\Delta
   Y_{\lambda_1}\,\bar s_{\lambda_1}\gm b_{\lambda_1} \sum_{j =
   u,d,s,c}Y_{\lambda_2 q_j}\,\bar q_{\lambda_2 j}\gamma_{\mu} q_{\lambda_2
   j}  + {\rm h.c.}\,;\nn
\\ \\
   \CH_{S} &=& \CH_{FU} +
       \frac{g^2_C\cot^2{\theta_C}}{2M^2_{V_8}}\sum_{\lambda_1, \lambda_2 =
       L,R}D^*_{\lambda_1 bs}D_{\lambda_1 bb}\, \bar s_{\lambda_1}\gm
       \frac{\lambda_A}{2} b_{\lambda_1} \nn \\
       & & ~~~~~~~~~~~~~~~~~~~~\times \Big(\sum_{j = d,s}\, 
       |D_{\lambda_2 bj}|^2 + \sum_{j = u,c} |U_{\lambda_2 tj}|^2 \Big)~\bar
       q_{\lambda_2 j}\gamma_{\mu} \frac{\lambda_A}{2} q_{\lambda_2 j} + {\rm
       h.c.}\,.\nn
\eea

The standard model contribution to a $\BD$-decay interaction is written as a
sum over a standard set of operators, each multiplied by the appropriate
Wilson coefficient. These coefficient functions are found at $M_W$ by
calculating the necessary QCD and electroweak (EW) penguin (loop) diagrams.
We rewrite the TC2 interactions in terms of the same set of operators by
Fierzing color octet products and using parity to relate matrix elements of
wrong chirality operators to the standard ones. The TC2 coefficient functions
involve combinations of hypercharges and $U(1)_1$ and $SU(3)_1$ couplings
rather than loop factors.
 
The standard operators have LL and LR chirality. Casting the RR and RL
operators from TC2 in the same color and charge structure as these, we obtain
the eight wrong chirality operators
\begin{eqnarray}\label{eq:wrong}
\hat Q'_3 &=& (\bar s b)_{V + A}\sum_q (\bar q q)_{V + A} \qquad\qquad\,\,
\hat Q'_5 = (\bar s b)_{V + A}\sum_q (\bar q q)_{V - A} \nonumber \\
\hat Q'_4 &=& (\bar s_{\alpha} b_{\beta})_{V + A}\sum_q (\bar q_{\beta}
              q_{\alpha})_{V + A} \qquad\,\,\,
\hat Q'_5 = (\bar s_{\alpha} b_{\beta})_{V + A}\sum_q (\bar q_{\beta}
            q_{\alpha})_{V - A} \nonumber \\ 
\hat Q'_7 &=& \frac{3}{2}(\bar s b)_{V + A}\sum_q e_q(\bar q q)_{V - A} \qquad\,\,\,\,
\hat Q'_9 = \frac{3}{2}(\bar s b)_{V + A}\sum_q e_q(\bar q q)_{V + A} \nonumber \\
\hat Q'_8 &=& \frac{3}{2}(\bar s_{\alpha} b_{\beta})_{V - A}\sum_q e_q(\bar
              q_{\beta} q_{\alpha})_{V + A} \quad
\hat Q'_{10} = \frac{3}{2}(\bar s_{\alpha} b_{\beta})_{V + A}\sum_q e_q(\bar
q_{\beta} q_{\alpha})_{V + A} \,,
\end{eqnarray}
where $e_q$ is the charge of quark $q$. The total TC2 contribution is the sum
of the standard and wrong chirality portions.\footnote{A factor of
  $G_F/\sqrt{2}$ has been taken out for easy comparison with the standard
  model contribution.}
\begin{equation}\label{eq:Heff}
   \CH_{eff, TC2}(\mu = M_{Z'}) = \frac{G_F}{\sqrt 2}\sum_{i = 3}^{10}
   \Big ( C_{i,TC2}(\mu)\hat Q_i(\mu) + C'_{i, TC2}(\mu)\hat
   Q'_i(\mu)\Big) \,.
\end{equation}  
Using parity For $\BD$-decays to a pair of pseudoscalars or a pesudoscalar
and a vector,
\begin{equation}\label{eq:parity}
       \langle PP |\hat Q_i'|\BD \rangle = -\langle PP| \hat Q_i|\BD \rangle \,,\quad
       \langle PV |\hat Q_i'|\BD \rangle  = +\langle PV| \hat Q_i|\BD \rangle \,.
\end{equation}
the effective Hamiltonian reduces to a sum over standard operators alone:
\begin{equation}\label{eq:7}
      \CH_{eff, TC2} = \frac{G_F}{\sqrt 2}\sum_{i = 3}^{10}(C_{i, TC2}
      \pm C'_{i, TC2})\hat Q_i\,,
\end{equation}
where the $+$ ($-$) sign is for $\BD \ra$ PV (PP).

The Wilson coefficient contributions $\tilde C_i = C_i \pm C'_i$ from coloron
and $Z'$ interactions are tabulated below. They are to be multiplied by the
mixing factor $D_{Lbb}D^*_{Lbs} \simeq V_{tb}V^*_{ts}$ and by $\sqrt{2}/(4
G_F)$. The coloron contributions are (up to smaller terms of $\CO(|D_{Lbd}|^2)$:
\begin{eqnarray}\label{eq:CIColoron}
  \tilde C_{3, V_8} & = & -\frac{g^2_C\cot^2{\theta_C}}{2
    M^2_C}\frac{|D_{Lbs}|^2}{9}\Big (1 \pm \chi^3
  e^{i\delta} \Big) \nonumber \\
  \tilde C_{4, V_8} & = &
  \frac{g^2_C\cot^2{\theta_C}}{2M^2_C}\frac{|D_{Lbs}|^2}{3}
  \Big(1\pm \chi^3 e^{i\delta} \Big) \nonumber \\
  \tilde C_{5, V_8} & = & -\frac{g^2_C\cot^2{\theta_C}}{2
    M^2_C}\frac{|D_{Lbs}|^2}{9}\Big (\chi^2 \pm \chi
  e^{i\delta} \Big)\nonumber \\
  \tilde C_{6, V_8} & = &
  \frac{g^2_C\cot^2{\theta_C}}{2M^2_C}\frac{|D_{Lbs}|^2}{3}
  \Big(\chi^2\pm \chi e^{i\delta} \Big) \nonumber \\
  \tilde C_{7, V_8} & = &  \frac{g^2_C\cot^2{\theta_C}}{2
    M^2_C}\frac{|D_{Lbs}|^2}{9}\Big (\chi^2 \pm \chi
  e^{i\delta} \Big) \nonumber \\
  \tilde C_{8, V_8} & = &
  -\frac{g^2_C\cot^2{\theta_C}}{2M^2_C}\frac{|D_{Lbs}|^2}{3}
  \Big(\chi^2\pm \chi e^{i\delta} \Big)\nonumber \\
  \tilde C_{9, V_8} & = & \frac{g^2_C\cot^2{\theta_C}}{2
    M^2_C}\frac{|D_{Lbs}|^2}{9}\Big (1 \pm \chi^3
  e^{i\delta} \Big) \nonumber \\
  \tilde C_{10, V_8} & =
  &-\frac{g^2_C\cot^2{\theta_C}}{2M^2_C}\frac{|D_{Lbs}|^2}{3}
  \Big(1\pm \chi^3 e^{i\delta} \Big) 
\end{eqnarray}
The $Z'$ contributions are:
\begin{eqnarray}\label{eq:CIZprime}
      \tilde C_{3, Z'} & = &\frac{g^2_Y\cot^2{\theta_Y}(\Delta
      Y_L)^2}{2M^2_{Z'}}\Big(\frac{Y_{Ld}}{\Delta Y_L} \pm \frac{2Y_{Rd} +
      Y_{Ru}}{3\Delta Y_L}\xi~\chi e^{i\delta}\Big) \nonumber \\
      \tilde C_{5, Z'} & = & \frac{g^2_Y\cot^2{\theta_Y}(\Delta
      Y_L)^2}{2M^2_{Z'}}\Big(\frac{2Y_{Rd} +
      Y_{Ru}}{3\Delta Y_L} \pm \frac{Y_{Ld}}{\Delta Y_L}\xi~\chi
      e^{i\delta}\Big)\nonumber \\
      \tilde C_{7, Z'} & = & \frac{g^2_Y\cot^2{\theta_Y}(\Delta
      Y_L)^2}{2M^2_{Z'}}\Big(\frac{2(Y_{Ru} -
      Y_{Rd})}{3\Delta Y_L}\Big) \nonumber \\
      \tilde C_{9, Z'} & = & \frac{g^2_Y\cot^2{\theta_Y}(\Delta
      Y_L)^2}{2M^2_{Z'}}\Big(\pm \frac{2(Y_{Ru} -
      Y_{Rd})}{3\Delta Y_L}~\xi~\chi~e^{i\delta}\Big)  
    \end{eqnarray}
The free parameters are $\xi$, $\chi$, and $\delta$, defined by
\begin{equation}\label{eq:xiepsil}
     \xi = \frac{\Delta Y_R}{\Delta Y_L}\,, \quad \chi = \Big|
     \frac{D_{Rbs}}{D_{Lbs}} \Big|\,, \quad \delta =
     \arg\Big(\frac{D_{Rbb}D^*_{Rbs}}{D_{Lbb}D^*_{Lbs}}\Big)\,.
\end{equation}
The $V_8$ and $Z'$ coefficients appear in the effective STC2 Hamiltonian,
while in FUTC2, the coloron coefficients are absent. The gluonic and
electroweak color singlet penguin operators $\hat Q_3, \hat Q_5, \hat Q_7$,
and $\hat Q_9$ receive contributions from both coloron and $Z'$ exchange,
while the color octet product penguin operators $\hat Q_4, \hat Q_6, \hat
Q_8$, and $\hat Q_{10}$ receive only coloron contributions. There is
no TC2 contribution to the standard tree level operators $\hat Q_1, \hat
Q_2$. In the definitions of $\tilde C_{i, TC2}$, we imposed the hypercharge
restrictions in Eq.~(\ref{eq:YGIM}). The nonstandard CP-violating terms are
proportional to $D_{Rbs}/D_{Lbs}$. For models with $D_R$ nearly
block-diagonal, the TC2 contributions are therefore coherent with the
standard ones and do not cause a $\sin{2\beta}$ discrepancy.

We have neglected the standard model contributions from penguin operators
with internal up or charm quarks. These contain CP conserving phases and are
consequently the source of direct CP violation within the standard model.
Since the TC2 interactions we have included are all tree level interactions,
the total SM + TC2 Hamiltonian will contain no direct CP violation. For
simplicity we have also neglected standard electromagnetic penguin and
chromomagnetic penguin operators. For reviews that include these operators
and their possible influence on CP violating $B$ decays see
Ref.\cite{Kramer:1994in,Grossman:1997gr}

To combine TC2 and standard-model effects, we need to run the TC2
contributions from $M_{Z'}$ down to $M_W$ using the renormalization group
equation (RGE). The RGE for the coefficient functions of the standard $\Delta
B = 1$ operators is known and has been calculated to several orders in
$\alpha_s$ (see Ref.~\cite{Buchalla:1996vs}). Loop-level gluon ($SU(3)_{1,2}$
gauge boson) effects can mix the operators, so the RGE for the coefficient
functions is a matrix equation.

An important approximation we make to obtain the RGE is to consider only QCD
renormalization effects. Electroweak contributions are negligible. But strong
$\uone$ and $\suone$ (in FUTC2) renormalizations are not. The former are very
model-dependent and their effect hard to predict. The latter are intractable
because of the strong $\alpha_C\cot\thc$ coupling, but it is not implausible
that they do not alter the pattern of operator mixing. The RGE is then
\begin{eqnarray}\label{eq:evo}
  C_{i, TC2}(M_{W}) &=& \sum_j \hat U(M_{Z'}, M_W)_{ij}\, C_{j, TC2}(M_{Z'})\,, \nn \\
  \hat U(M_{Z'}, M_W)_{ij} & = &
          \exp\Bigg[\int_{g_1(M_W)}^{g_1(M_{Z'})}dx
          \frac{\gamma^T(x)_{ij}}{\beta(x)} \Bigg] \,.
\end{eqnarray} 
Here $\gamma^T(x)$ is the transposed anomalous dimension matrix, and
$\beta(x)$ the QCD beta function. In our calculations, we used the
$\CO(\alpha_C)$ $\beta$ and $\gamma$-functions~\cite{Buchalla:1996vs} and
included only standard model particles\footnote{We adjusted the
  renormalization and subtraction scheme in~\cite{Buchalla:1996vs} to be
  consistent with the current results for $M_W$, $m_t$, and $\alpha_s$}.  After
running the TC2 effects, we have the total standard plus TC2 Hamiltonian at
$M_W$.
\begin{equation}\label{eq:CHeff}
  \CH_{\rm eff}(M_W) = \frac{G_F}{\sqrt 2}\Big [V_{ub}V^*_{us}(C_1\hat Q^u_1 +
  C_2 \hat Q^u_2) + V_{cb}V^*_{cs}(C_1\hat Q^c_1 +
  C_2 \hat Q^c_2) - V_{tb}V^*_{ts}\sum_{i = 3}^{10}(C_{i,SM} -
  \tilde C_{i,TC2})\hat Q_i\Big ] \,.
\end{equation}

Finally, we run the standard-model plus TC2 Wilson coefficients from $M_W$ down to
the desired energy $\mu$ (here, $m_b$):
\be\label{eq:Cimu}
C_{i}(\mu) = \sum_j \hat U(M_W, \mu)_{ij}~C_j(M_W) \,,
\ee
where
\be\label{CiMw}
C_{i}(M_W) = C_{i, SM}(M_W) + \sum_j\hat U(M_{Z'},M_W)_{ij}~C_{j, TC2}(M_{Z'})\,.
\ee
The evolution matrix in $C_i(\mu)$ has the same form as in Eq.~(\ref{eq:evo})
except that limits involve the QCD coupling $g_C$ rather than the $\suone$
coupling. The anomalous dimension matrix and the beta function in
Eq.~(\ref{eq:Cimu}) include only five quark flavors. The resulting
Hamiltonian, with possible new CP-violating phases from $D_R$, is
\be\label{eq:HAM}
\CH_{\rm eff}(b \ra s\bar q q) = 
  \frac{G_F}{\sqrt 2}~V_{tb}V^*_{ts}~\sum_i \tilde D_{i}(\mu)\hat Q_i(\mu)
\ee
where
\be\label{eq:Dtilde}
\tilde D_i(\mu) = \sum_j\hat U(M_W, \mu)_{ij}\Big (C_{j, SM}(M_W)
  + \sum_k \hat U(M_{Z'}, M_W)_{jk}~C_{k, TC2}(M_{Z'})\Big )\,.
\ee

To apply this Hamiltonian to a particular $b \ra s \bar q q$ process, we
evaluate the amplitude using the factorization
approximation~\cite{Ali:1998eb, Ali:1998nh}. There the operators are split
into two subcurrents:
\begin{equation}\label{eq:fact}
  \langle h_1h_2|\hat Q_i = j_{1\mu}j^{\mu}_2 |\BD \rangle \cong \langle
  h_1|j_{1\mu}|0\rangle \langle h_2|j^{\mu}_2|\BD \rangle + \langle
  h_2|j_{1\mu}|0 \rangle \langle h_1|j^{\mu}_2|\BD \rangle\,,
\end{equation}
where we ignored annihilation terms such as $\langle
h_1h_2|j_{1\mu}|0\rangle \langle 0|j_{2\mu}|B \rangle$. The $\langle
h_i|j_\mu|\BD \rangle$ portion is a form factor that can be measured in a
semileptonic decay, while $\langle h_j|j_\mu|0\rangle$ is a measurable decay
constant. The values of the form factors and decay constants we used in our
calculations can be found in Ref.~\cite{Ali:1998eb}. Different operators with
different chiral and color prefactors will
contribute depending on the particular decay process. For example:
\begin{eqnarray}\label{eq:BtophiKs}
  A(B \ra \phi K_S) &\propto& \langle\phi |(\bar ss)_{V -A}|0\rangle
  \langle K_S|(\bar s b)_{V - A} |B\rangle \Big[(\tilde D_3 + \tilde
  D_4 - \half \tilde D_9 - \half \tilde D_{10})\Big( \frac{1}{N_{QCD}} +
  1 \Big) \nn\\
  &+& \tilde D_5 - \half \tilde D_7 + \frac{1}{N_{QCD}}(\tilde D_6
  - \half \tilde D_8) \Big] \,.
\end{eqnarray}
The factorized amplitude includes only the contribution from color singlet
operators. To compensate for the contributions from other operators, the
number of color $N_C$ is treated as a parameter\footnote{For all calculations
  we use $N_C = 3$}. Because of the possible dependency on the effective
number of colors (and other more technical reasons), we caution that the
factorization approximation is not a good approximation for all modes.
Factorization for the decay modes $\eta' K_S, \pi K_S$, and to some extent
$\phi K_S$, is considered to be reliable~\cite{Ali:1998nh}.\footnote{Due to tree
  dominance of the $J/\psi$ mode, factorization is not carried out in the
  same manner as the other modes, but it is still reliable.}  After the
amplitude is factored, it is useful to separate its real and imaginary parts:
\begin{eqnarray}\label{eq:facttwo}
  &&A(B \ra f) = V_{CKM}(\CX + i\CY)\,,\nn\\
  &&\CX = \sum_i {\rm Re}(a_i\tilde D_i) \,,\quad \CY = \sum_i {\rm
  Im}(a_i\tilde D_i)\,.
\end{eqnarray}
where the $a_i$ are numerical factors multiplying the coefficient functions
($\tilde D_i$) in the factorized amplitude and $V_{CKM}$ is the standard
model CKM factor for the process. {\em In models with block-diagonal $D_R$,
  $\CY \cong 0$ for the $\BD \ra X K_S$ decays we consider and all correct
  measurements return the same value of $\sin 2\beta$.}

\section*{6. The extraction of $\sin{2\beta}_{\rm eff}$}

With the Hamiltonian in Eq.~(\ref{eq:HAM}) renormalized and factorized at
$m_b$, we proceed with the standard CP formalism described in many
review papers~\cite{Buras:2001pn, Fleischer:2002ys}. The asymmetry we
are interested in for comparison to $\sin{2\beta}$ involves interference
between the $\BDbar$--$\BD$ mixing phase $\phi_M$ and the decay phase
$\phi_D$. It is defined by
\begin{equation}\label{eq:amd}
  a_{MD}(t)  =  \frac{\Gamma(B^0_{phys}(t)\ra f) -
    \Gamma(\bar{B}^0_{phys}(t)\ra f)}{\Gamma(B^0_{phys}(t)\ra f) +
    \Gamma(\bar{B}^0_{phys}(t)\ra f)} \,.
\end{equation} 
The state $B^0_{phys}$ is a meson that started at production time $t = 0$ as
a $\BD$ but contains both $\BD$ and $\BDbar$ at later times. The CP
asymmetry is described in terms of the phase-convention-independent parameter
$\lcp$:
\begin{equation}\label{eq:lambdaCP}
  \lambda_{CP}  = \Big(\frac{q}{p} \Big)\frac{A(\bar{B}^0 \ra 
    \bar{f})}{A(B^0 \ra f)}  =  \eta \Big( \frac{q}{p}
  \Big)\frac{A(\bar{B}^0 \ra f)}{A(B^0 \ra f)}\,
\end{equation}
where $\eta$ is the CP eigenvalue of the final state.\footnote{Therefore this
  formalism only applies to final states that are CP eigenstates.} The $q/p$
factor comes from $B$ mixing and describes the proportion of $\BD$ to
$\BDbar$ in the mass eigenstates. It is a pure phase, but with the addition
of TC2 effects that phase may no longer be $\beta = \arg(V^*_{td})$, so we
write
\begin{equation}\label{eq:qoverp}
  \Big(\frac{q}{p}\Big)_{TC2} = e^{-2i\phi_M} \,.
\end{equation}
We can write Eqs.~(\ref{eq:DelMFU},\ref{eq:DelMS}) as the standard model
phase $\arg(2 V^*_{td})$ times some complex number. The phase of this complex
number is the nonstandard mixing phase. Since only the TC2$_{LR}$ and TC2$_{RR}$
contributions contain phases different from $\arg(2 V^*_{td})$,
the mixing phase is
\begin{equation}\label{eq:phiM}
  \phi_M = \arg(V^*_{td}) + \arctan{\Big[\frac{{\rm Im}({\rm TC2_{RR} +
  TC2_{LR}})}{{\rm Re}{\rm (SM + TC2)}}\Big]}\,.
\end{equation}

In the standard model, the amplitude ratio $\bar A/A$ for the decay modes
$\BD \ra XK_S$ has unit magnitude and no imaginary part (to within 4\%).

 Since our standard
plus TC2 Hamiltonian contains no sources of direct CP violation, the
magnitude of the amplitude ratio will not change. The addition of TC2 effects
therefore alters only the phase of $\bar A/A$:
\be \label{eq:randphi}
\phi_D = \half \arctan{\Big[\frac{-2\CX\CY}{\CX^2 - \CY^2}\Big]} \,,
\ee
and $\CX$ and $\CY$ are the real and imaginary parts of $A$ (see
Eq.~(\ref{eq:facttwo})). It is possible to obtain a $\sin{2\beta}$ discrepancy
with an additional decay phase but no additional mixing phase.The value of
$\phi_D$ depends on the final state, since the operators that contribute to a
decay and their relative strength depend on the decay mode
and are determined by the factorization. If there is no new CP-violating
decay phase, then $\CY = \phi_D = 0$.

Expressed in terms of $\lcp$, the asymmetry at time $t$ is
\begin{equation}\label{eq:amdtwo}
  a_{MD} = \frac{(1-|\lcp|^2)\cos{(\Delta Mt)}  -
    2{\rm Im}(\lcp)\sin{(\Delta Mt)}}{(1 + |\lcp|^2)} 
\end{equation} 
The term we are interested in is the one proportional to ${\rm Im}(\lcp)$.
In the standard model, $\phi_D = 0$ so that ${\rm Im}(\lcp) = -\sin{2\beta}$.
In TC2 models, the ${\rm Im}(\lcp)$ term becomes
%
\begin{equation}\label{eq:sineff}
  \sin{2\beta}_{\rm eff} = \sin{(2(\phi_M - \phi_D))}.
\end{equation}
Any discrepancy among the various decay modes is therefore due to differing decay
phases.

\section*{7. Comparisons with Experiment}

\subsection*{7.1 $\sin{2\beta}_{\rm eff}$ in $\BD \ra X K_S$ Decays}

Using this formalism, we calculated $\sin{2\beta}_{\rm eff}$ for the decays
$\BD \ra J/\psi K_S$, $\phi K_S$, $\eta' K_S$ and $\pi K_S$. The current
experimental values are recorded again here. They are unsettled, but seem to
show a discrepancy, especially between $J/\psi K_S$ and the Belle measurement
of $\phi K_S$ and, possibly, $\eta' K_S$:
\bea\label{eq:XKsDatatwo}
\sin{2\beta}_{J/\psi K_S} &=& +0.72\pm
0.05\qquad\qquad\,\,{\cite{Browder:2003ii}}\nn\\
\sin{2\beta}_{\phi K_S} &=& +0.47\pm 0.34\quad({\rm Babar}\,\,
{\cite{Aubert:2004dy}})\nn\\
\sin{2\beta}_{\phi K_S} &=& +0.06\pm 0.33\quad({\rm Belle}\,\,\,\,\,
{\cite{Abe:2004xp}})\\
\sin{2\beta}_{\eta' K_S} &=& +0.27\pm
0.21\,\,\qquad\qquad{\cite{Aubert:2003bq}}\nn\\
\sin{2\beta}_{\pi K_S} &=& +0.48^{+0.38}_{-0.47}\pm0.11
\quad\quad\,\,\,{\cite{Aubert:2004xf}}\nn
\eea

In Mass Model~1, $|D_{Rbs}| \ll |D_{Lbs}|$ and $|D_{Rbd}| \ll |D_{Lbd}|$, so
$\sin{2\beta}_{\rm eff}$ is the same for all modes, with $\beta_{\rm eff} =
\arg(V^*_{td})$ = 0.516 (see Appendix~B). This and the other models were not
tuned to give $\arg(V^*_{td}) = \beta_{J/\psi K_S}$, but it would not be
difficult to do so. Nor did we attempt to match the experimental $V_{cb}
\simeq 0.04$, so that the $|V_{ts}| \simeq |V_{cb}|$ we use below to
determine the standard model's contribution to $\Delta M_{B_s}$ is too small.

There can be a sizable differences in the values of $\sin{2\beta}_{\rm eff}$
for $J/\psi K_S$ and the other modes in Mass Model~2. The discrepancies
$|\sin{2\beta}_{X K_S} - \sin{2\beta}_{J/\psi K_S}|$ are plotted in
Figs.~\ref{fig:CPVB_fig_1}--\ref{fig:CPVB_fig_6} for both TC2 variants as a
function of the parameters $Y_{Ru}$ and $Y_{Rd}$.\footnote{The sharp features
  in these figures are caused by the arctangent in the expression for
  $\phi_D$ becoming large whenever the argument in the denominator of
  (\ref{eq:randphi}) approaches zero.} We again assumed the $Z'$ and $V_8$
couplings are half-critical (Eq.~(\ref{eq:halfcrit})) and that $\Delta Y_L^2
= Y_{Lt}Y_{Rt}$. To maximize the TC2 contribution, we used the lowest bounds
on $M_{Z'} = M_{V_8}$ found in Eq.~(\ref{eq:MVZlimits}). (For a similar
analysis that contains only the coloron contribution, see
Ref.~\cite{Burdman:2003nt}). The parameters $\Delta Y_L = -0.5$ and $Y_{Lq} =
1$ were chosen to avoid light quark condensation. Only negative hypercharges,
$Y_{Ru}$ and $Y_{Rd}$, are included in the FUTC2 variant because of
condensation constraints. Clearly, large discrepancies are possible in both
TC2 variants. In STC2, $|Y_{Ru}|$ and $|Y_{Rd}| \simle 2$ are sufficient
to produce the central values of the discrepancies. Because of the larger
$M_{Z',\,V_8}$ in FUTC2, somewhat larger $|Y_{Rd}|$ are needed to produce the
discrepancies for $\phi K_S$ and $\eta' K_S$. As we have discussed, large
hypercharges and a strong $\uone$ coupling are problematic because they make
the $\uone$ Landau pole occur at an uncomfortably low scale.

In Mass Model~3, the only appreciable difference from $\sin 2\beta_{\rm eff}$
for $J/\psi K_S$ for moderate hypercharges occurs for $\phi K_S$ in the STC2
variant of the model. This case is shown in Fig.~\ref{fig:CPVB_fig_7}. Large
discrepancies in the FUTC2 case require even larger hypercharges than in
Model~2. The discrepancies are generally much smaller than in Model~2 because
$M_{Z',V_8}$ are about twice as large in Model~3.


%
 \begin{figure}[t]
   \begin{center}
     \includegraphics[width=3.0in, height = 3.0in, angle=0]{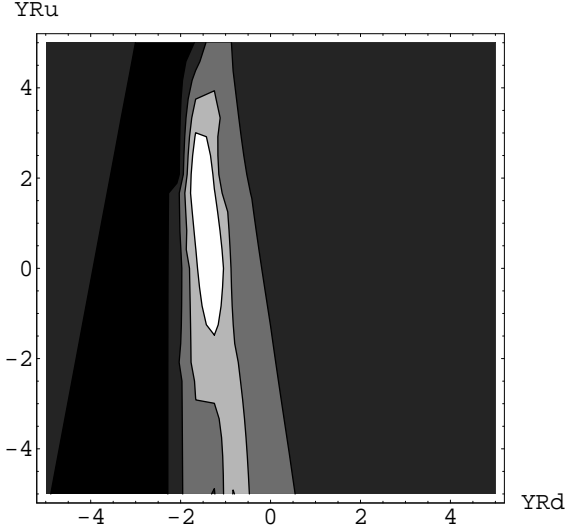}
     \caption{Discrepancy  $|\stwobeta_{\phi K_S} - \stwobeta_{J/\psi K_S}|$
       in the STC2 variant of Mass Model~2 as a function of $Y_{Rd}$ and $Y_{Ru}$.
       $|\stwobeta_{\phi K_S} - \stwobeta_{J/\psi K_S}| \ge 1$
       (black),~$>0.75$,~$>0.5$,~$>0.25$,~$<0.25$~(white). }
     \label{fig:CPVB_fig_1}
   \end{center}
 \end{figure}
 \begin{figure}[!h]
   \begin{center}
     \includegraphics[width=3.0in, height = 3.0in, angle=0]{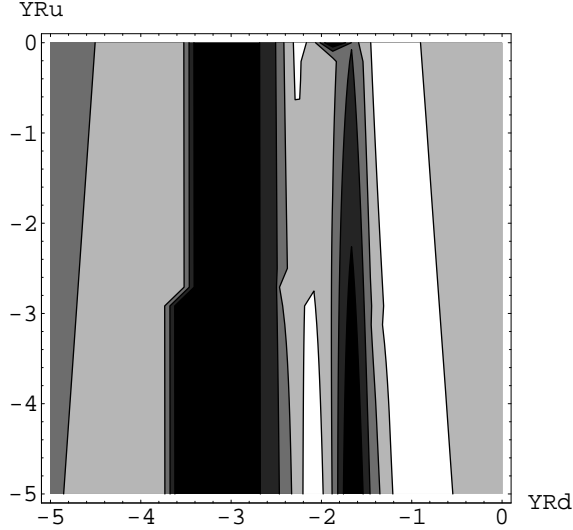}
     \caption{Discrepancy $|\stwobeta_{\phi K_S} - \stwobeta_{J/\psi K_S}|$ in
       the FUTC2 variant of Model~2, as a function of $Y_{Rd}$ and $Y_{Ru}$.
       Greyscale is the same as in Fig.~\ref{fig:CPVB_fig_1}.}
     \label{fig:CPVB_fig_2}
   \end{center}
 \end{figure}
\begin{figure}[t]
  \begin{center}
    \includegraphics[width=3.0in,height=3.0in, angle=0]{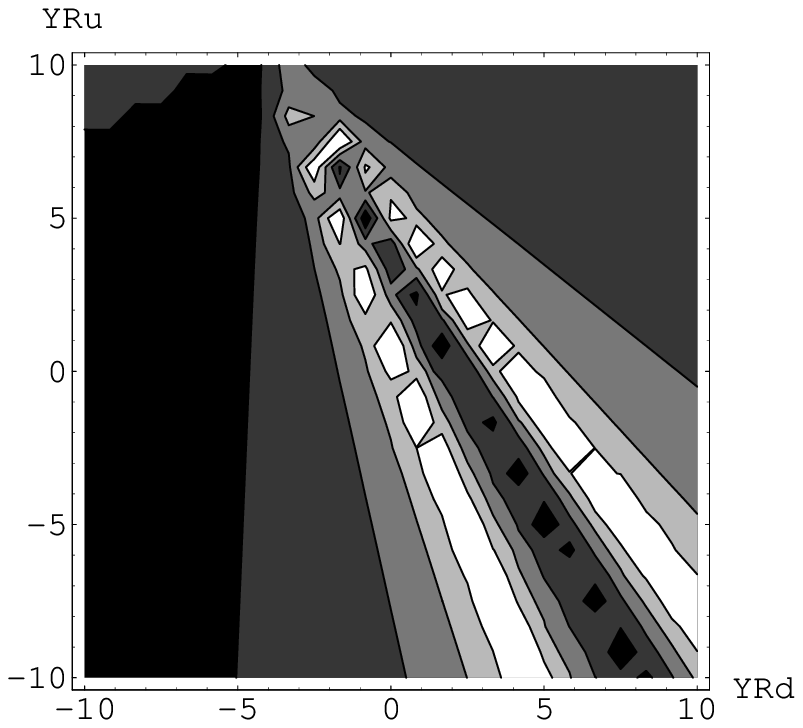}
    \caption{$|\stwobeta_{\eta' K_S} - \stwobeta_{J/\psi K_S}|$ as a
      function of $Y_{Rd}$ and $Y_{Ru}$ in the STC2 variant of Model~2.
      Greyscale is the same as in Fig.~\ref{fig:CPVB_fig_1}.}
    \label{fig:CPVB_fig_3}
  \end{center}
\end{figure}
\begin{figure}[!h]
  \begin{center}
    \includegraphics[width=3.0in,height=3.0in, angle=0]{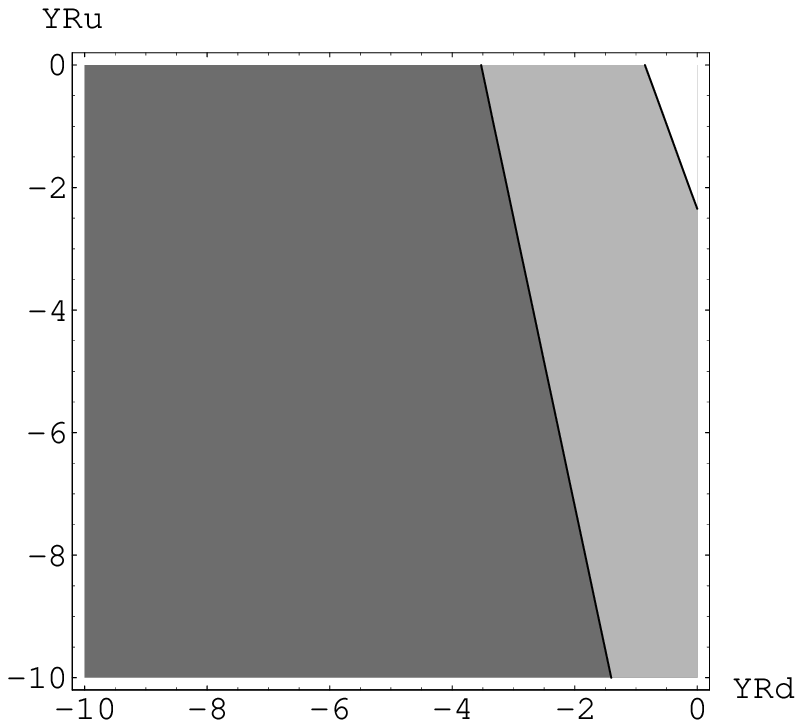}
    \caption{$|\stwobeta_{\eta' K_S} - \stwobeta_{J/\psi K_S}|$ as a
      function of $Y_{Rd}$ and $Y_{Ru}$ in the FUTC2 variant of Model~2.
      Greyscale is the same as in Fig.~\ref{fig:CPVB_fig_1}.}
    \label{fig:CPVB_fig_4}
  \end{center}
\end{figure}
\begin{figure}[!h]
  \begin{center}
    \includegraphics[width=3.0in,height=3.0in, angle=0]{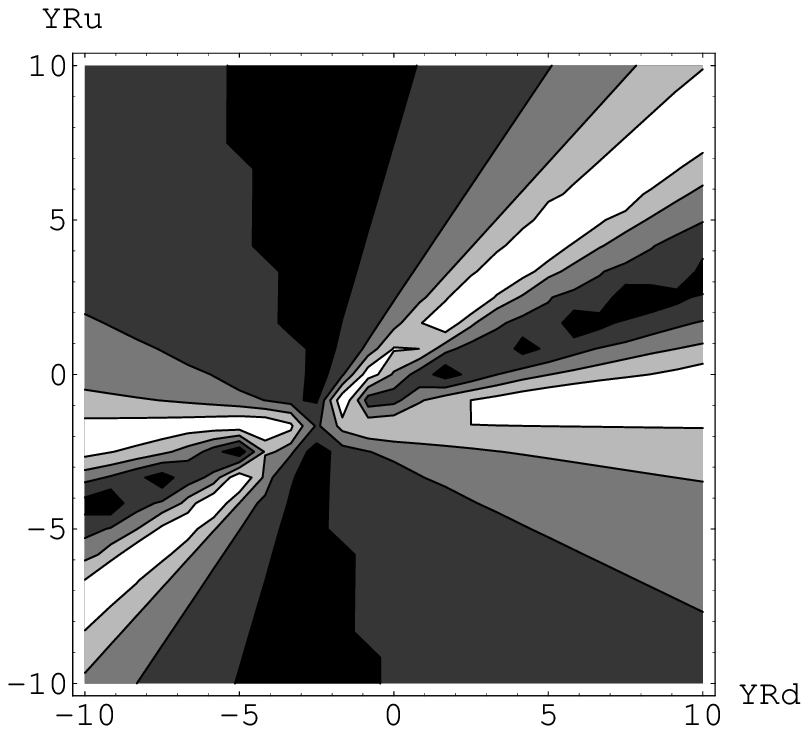}
    \caption{$|\stwobeta_{\pi K_S} - \stwobeta_{J/\psi K_S}|$ as a
      function of $Y_{Rd}$ and $Y_{Ru}$ in the STC2 variant of Model~2.
      Greyscale is the same as in Fig.~\ref{fig:CPVB_fig_1}.}
    \label{fig:CPVB_fig_5}
  \end{center}
\end{figure}
\begin{figure}[!h]
  \begin{center}
    \includegraphics[width=3.0in,height=3.0in, angle=0]{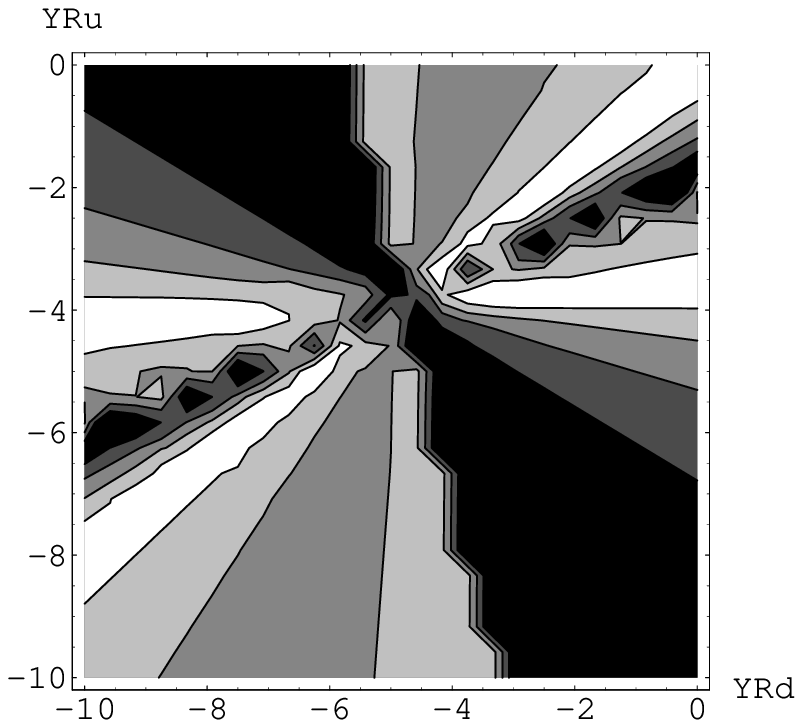}
    \caption{$|\stwobeta_{\pi K_S} - \stwobeta_{J/\psi K_S}|$ as a
      function of $Y_{Rd}$ and $Y_{Ru}$ in the FUTC2 variant of Model~2.
      Greyscale is the same as in Fig.~\ref{fig:CPVB_fig_1}.}
    \label{fig:CPVB_fig_6}
  \end{center}
\end{figure}
\begin{figure}[!h]
  \begin{center}
    \includegraphics[width=3.0in,height=3.0in, angle=0]{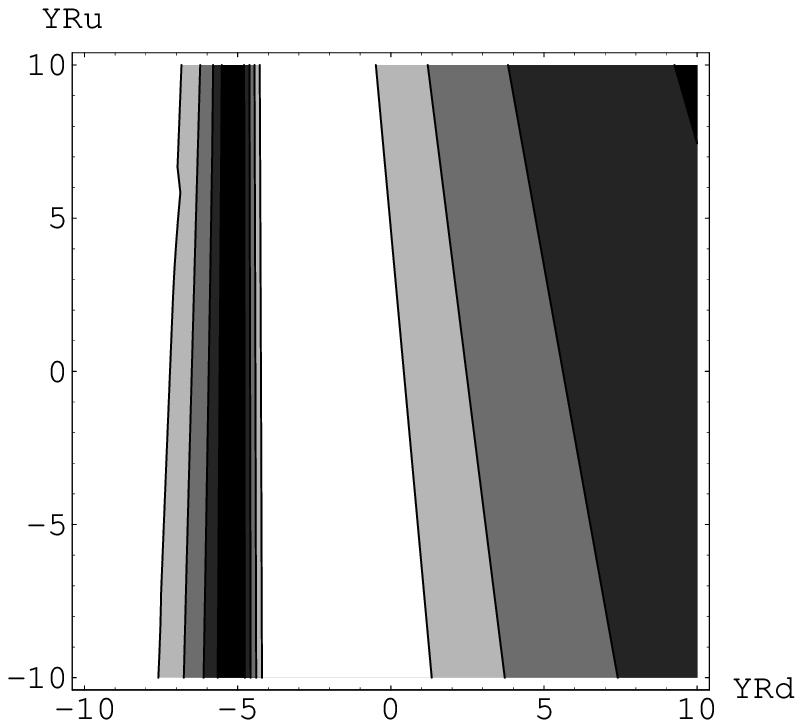}
    \caption{$|\stwobeta_{\phi K_S} - \stwobeta_{J/\psi K_S}|$ as a
      function of $Y_{Rd}$ and $Y_{Ru}$ in the STC2 variant of Mass Model~3.
      Greyscale is the same as in Fig.~\ref{fig:CPVB_fig_1}.}
    \label{fig:CPVB_fig_7}
  \end{center}
\end{figure}

\subsection*{7.2 Other TC2 Effects: $\Delta M_{B_s}$ and $\epsilon'/\epsilon$}

With $M_{Z',V_8}$ and $\Delta Y_R/\Delta Y_L$ deduced from $\Delta M_{B_d}$,
we can predict $\Delta M_{B_s}$. Its experimental lower bound is
$8.622\times10^{-9} \,\mev$~\cite{Eidelman:2004wy}. The formalism for
calculating the mass difference is similar to that used to obtain
Eqs.~(\ref{eq:DelMFU},\ref{eq:DelMS}) for $M_{12}(B_d)$. As before, ETC
contributions are negligible because we assumed they are
generation-conserving up to mass mixing. For TC2, the strange quark mass
$m_s$ replaces $m_d$, the CKM factor changes to $V_{tb}V^*_{ts}$, and
$D^*_{L,R\,bd}$ to $D^*_{L,R\,bs}$. Additionally, the $SU(3)$-breaking
difference between $F_{B_s}\sqrt{\hat B_s}$ and $F_{B_d}\sqrt{\hat B_d}$ is
included. As was the case for $\Delta M_{B_d}$, all TC2 contributions occur
at tree level, leaving $\Gamma_{12}(B_s)$ unchanged from its standard model
prediction.

The mass difference $\Delta M_{B_s}$ was calculated in both TC2 variants of
Mass Models~1--3, using the TC2 gauge boson masses in
Eq.~(\ref{eq:MVZlimits}). We obtained for $\Delta M_{B_s}$ and $x_s = \Delta
M_{B_s}/\Gamma_{B_s}$:
\begin{equation}\label{eq:DelMBs}
\begin{array}{cccc} \underline{{\rm MASS}\ {\rm MODEL}} & 
\underline{{\rm TC2}\ {\rm MODEL}} &  \underline{\Delta M_{B_s}(\mev)\times 10^{-8}} &
\quad \underline{x_s}\\
         1 & {\rm STC2} &  1.24   & \quad 27 \\
         2 & {\rm STC2} &  13.2   & \quad 293 \\
         3 & {\rm STC2} &  1.03   & \quad 23 \\
           &      &  \\
         1 & {\rm FUTC2} &  1.24  & \quad 27 \\
         2 & {\rm FUTC2} &  3.99  & \quad 88 \\
         3 & {\rm FUTC2} &  1.01  & \quad 22 \\
       \end{array}
\end{equation}

The agreement between the two cases in Model~1 is fortuitous. The values
would be different had we used values of $M_{V_8} = M_{Z'}$ larger than the
lower bounds. The standard model contribution to $\Delta M_{B_s}$ is
$8.96\times 10^{-9}\, \mev$, slightly larger than the current experimental
lower bound.  The results for Mass Model~2 are quite different. The standard
model contribution is smaller in this case, $(\Delta M_{B_s})_{SM} =
4.48\times 10^{-9}\,\mev$, because $|V_{ts}| \cong |D_{Lbs} D^*_{Lbb}|$ is
smaller than it is in Model~1. Had we tuned our CKM matrix to give the value
of $|V_{ts}| \cong |V_{cb}|$ in Ref.~\cite{Eidelman:2004wy}, it would double.
The TC2 contribution is much larger here than that from the standard model
because $M_{Z',V_8}$ are 20--30\% what they are in Model~1. The smaller
values of $\Delta M_{B_s}$ in Model~3 are due mainly to the larger TC2 boson
masses.  The standard model contribution in this case is the same as in
Model~2 since Models~2 and~3 have almost identical CKM matrices. For all of
these mass models, TC2 contributions have a much larger effect on $\Delta
M_{B_s}$ than they do on $\Delta M_{B_d}$ because the CKM factor is larger:
$|V_{tb}V^*_{ts}| \cong |D_{Lbb}D^*_{Lbs}| \sim 5 |V_{tb}V^*_{td}| \cong 5
|D_{Lbb}D^*_{Lbd}|$. In short, our TC2 mass models can accommodate values of
$x_s$ ranging from~1 to~15 times the current experimental lower bound.

Finally, we calculated the quantity ${\rm Re}(\epsilon'/\epsilon)$ measuring
the ratio of direct to indirect CP violation in $K^0 \to \pi\pi$ decays. It
depends on the relative size and phases of the $\Delta I = \thhalf$ and
$\Delta I = \half$ amplitudes. Tree level and EW penguin operators contribute
to both isospin portions, while gluonic penguin operators contribute only to
$\Delta I = \half $. The world-average experimental value is ${\rm
  Re}(\epsilon'/\epsilon) = 16.6\pm 1.6 \times 10^{-4}$~\cite{Buras:2003zz}
while the latest standard model predictions are in the range $5$--$30\times
10^{-4}$~\cite{Buras:2001au}.
  

The TC2 contributions to $K^0 \to \pi \pi$ are incorporated following the
procedure of Sect.~5.\footnote{The ETC contributions to $K^0 \to \pi \pi$,
  calculated at the ETC scale, are highly suppressed by the large ETC gauge
  boson masses and by mixing angles. (They were first estimated in 2000 by
  G.~Burdman (unpublished), and we concur with him.) Running effects may
  enhance them, but not enough to make them comparable to the standard model
  contributions --- except, possibly, in the case of FUTC2 where quarks have
  strong $\suone$ interactions. As for Eq.~(\ref{eq:ratio}), we assume that
  quark anomalous dimensions are not large in FUTC2.} They are written in
terms of standard $\Delta S = 1$ operators, run down to $M_W$ using the RGE,
and combined with the standard model contributions. The TC2 $\Delta S = 1$
Wilson coefficients before running are similar to the $\Delta B = 1$
coefficients in Eqs.~(\ref{eq:CIColoron},\ref{eq:CIZprime}), with $m_s$
replacing $m_b$, and $D_{\lambda bs}D^*_{\lambda bd}$ replacing $D_{\lambda
  bb}D^*_{\lambda bd}$. For this kaon system observable, the standard model
plus TC2 Hamiltonian must be evaluated near $1\,\gev$.  Running down to $m_b$
is carried out in the same way as before. To evolve from $m_b$ to $1\,\gev$,
we must remove the bottom and charm quarks at the appropriate energies. This
requires successively mapping a five quark theory onto an effective four
quark theory, then the four quark theory onto a three quark one; see
Ref.~\cite{Buchalla:1996vs}.

Once the effective Hamiltonian at $1\,\gev$ is obtained, the expression for
${\rm Re}(\epsilon'/\epsilon)$ including TC2 contributions can be obtained by
generalizing the expressions given in (see
Refs.~\cite{Buchalla:1996vs,Buras:2003zz})
\begin{equation}\label{eq:epsprime}
  {\rm Re \Big (\frac{\epsilon'}{\epsilon}\Big)} = {\rm
  Im}\Big[V_{ts}V^*_{td}(P_{tot}^{(1/2)} - P_{tot}^{(3/2)})\Big]\,.
\end{equation}
Here, $P_{tot}^{(\Delta I)}$ contains the matrix elements of the standard
model plus the $|\Delta I| = \half$ and~$\thhalf$ contributions from
TC2.\footnote{The $|\Delta I| = \half$ and~$\thhalf$ matrix elements were
  taken from Ref.~\cite{Buchalla:1996vs,Buras:2003zz}, except for $\langle
  \hat Q_6 \rangle_0$ and $\langle \hat Q_8 \rangle_2$, where we used the
  large-$N_C$ lattice results given in Ref.~\cite{Buras:2001pn}.} The
additional phases in the TC2 contributions may make these matrix elements
complex. We use as inputs the experimental values $\epsilon = (2.271\pm
0.017)\times 10^{-3}~\exp(i\pi/4)$, the $\Delta I = \half$ amplitude ${\rm
  Re}A_0 = 3.33\times 10^{-7}\,\gev$, and the ratio ${\rm Re}A_2/{\rm Re}A_0
= 0.045$.  Only $\epsilon$ receives appreciable ETC and TC2 contributions
and, as we noted earlier, it is not a stringent constraint on TC2. The other
two inputs (${\rm Re}A_0$ and ${\rm Re}A_2/{\rm Re}A_0$) have mainly standard
model contributions. Using the same values for $\Delta Y_L$ and
$Y_{Ld}$ as for $\stwobeta$, we calculated ${\rm Re}(\epsilon'/\epsilon)$ as
a function of the $Y_{Rd}$ and $Y_{Ru}$ hypercharges. In the table in
Eq.~(\ref{eq:epsprimetable}), we list the standard model contribution ---
whose phase is contained in ${\rm Im}(V_{ts}V^*_{td})$, the TC2 contribution
involving $Y_{Ld}$ --- whose phase is the same, and the TC2 contribution from
$D_R$, which is proportional to $Y_{Rd} - Y_{Ru}$, i.e., ${\rm
  Re}(\epsilon'/\epsilon) = {\rm SM} + {\rm TC2}(Y_{Ld}) + {\rm TC2}
\times(Y_{Rd} - Y_{Ru})$. The $V_8$ contribution is negligible compared to
the standard one.
\begin{equation}\label{eq:epsprimetable}
\begin{array}{ccccc} \underline{{\rm MASS}\ {\rm MODEL}} & 
\underline{{\rm TC2}\ {\rm MODEL}} &  \underline{{\rm SM}\times 10^{-4}} &
\underline{{\rm TC2}(Y_{Ld})\times 10^{-4}} &
\underline{{\rm TC2}\times(Y_{Rd} - Y_{Ru})10^{-4}}\\
1 & {\rm STC2} &  13.3 & -0.18  & 27\\
2 & {\rm STC2} &  10.3 & -10.8  & 422  \\
3 & {\rm STC2} &  10.1 & -0.50 & 100 \\
  &     \\
1 & {\rm FUTC2} &  13.3 & -0.23  & 34\\
2 & {\rm FUTC2} &  10.3 & -5.52  & 217\\
3 & {\rm FUTC2} &  10.1 & -0.53  & 107\\
\end{array}
\end{equation}

\begin{figure}[!h]
  \begin{center}
    \includegraphics[width=3.0in, height=3.0in, angle=0]{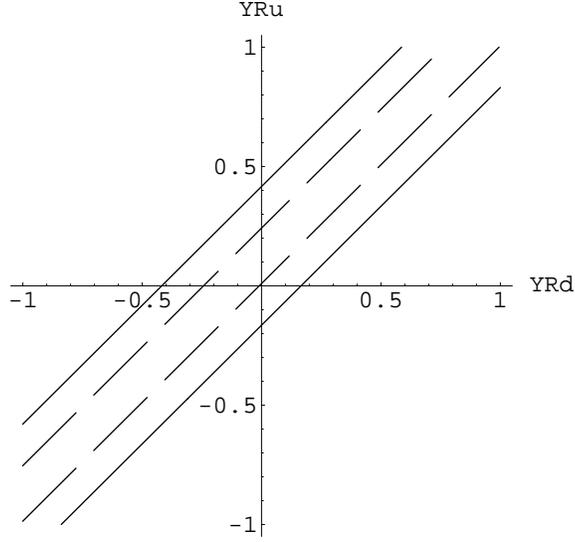}
    \caption{${\rm Re(\epsilon'/\epsilon)}$ as a function of $Y_{Rd}$ and
      $Y_{Ru}$ in the STC2 variant of Model~1. Solid lines indicate $\pm 5\
      \sigma$, dashed lines $\pm 2\ \sigma$ from the central experimental
      value of $16.6\times 10^{-4}$.}
    \label{fig:CPVB_fig_8}
  \end{center}
\end{figure}
In Figs.~\ref{fig:CPVB_fig_8}--\ref{fig:CPVB_fig_11} we plot, for the
Models~1 and~2, bands in the $Y_{Rd}$--$Y_{Ru}$ plan corresponding to two and
five sigma spreads from the central experimental value of ${\rm
  Re}(\epsilon'/\epsilon)$. For Models~1 and 3 the plots of ${\rm
  Re(\epsilon'/\epsilon)}$ in FUTC2 and in STC2 are nearly identical, a
result of both TC2 variants having approximately the same minimum $Z'$ mass.
It is clear from the table that $Y_{Rd} - Y_{Ru}$ must be close to zero (very
close for Model~2) and slightly negative. The two hypercharges must be so
close in Model~2 that they would be a strain on building a complete model
with its mass matrix.
%
%
%
\begin{figure}[!h]
  \begin{center}
    \includegraphics[width=3.0in, height=3.0in, angle=0]{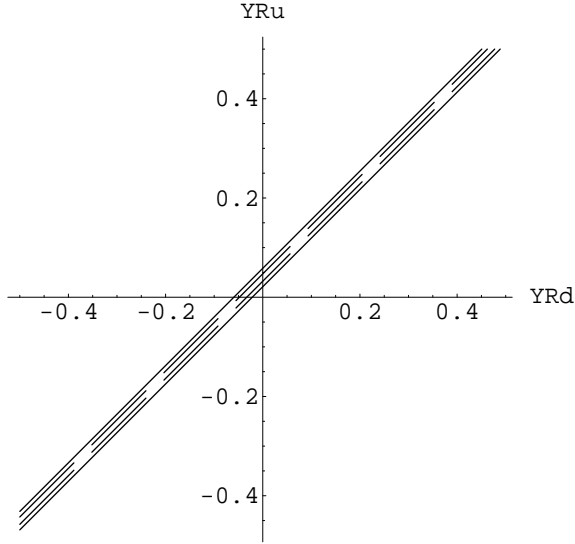}
    \caption{${\rm Re(\epsilon'/\epsilon)}$ as a function of $Y_{Rd}$ and
      $Y_{Ru}$ in the STC2 variant of Model~2. The $2\sigma$ and $5\sigma$
      bands are the same as in Fig.~\ref{fig:CPVB_fig_8}.}
    \label{fig:CPVB_fig_10}
  \end{center}
\end{figure}
\begin{figure}[!h]
  \begin{center}
    \includegraphics[width=3.0in,height=3.0in,angle=0]{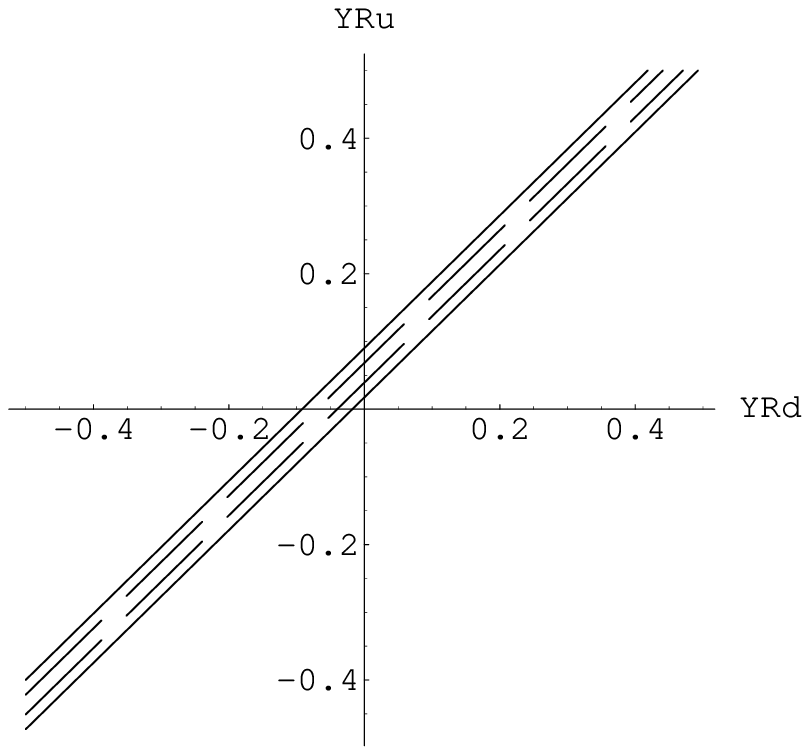}
    \caption{${\rm Re}(\epsilon'/\epsilon)$ as a function of $Y_{Rd}$ and
      $Y_{Ru}$ in the FUTC2 variant of Model~2. The $2\sigma$ and $5\sigma$
      bands are the same as in
      Fig.~\ref{fig:CPVB_fig_8}.}
    \label{fig:CPVB_fig_11}
  \end{center}
\end{figure}
\begin{figure}[!h]
  \begin{center}
    \includegraphics[width=3.0in, height=3.0in, angle=0]{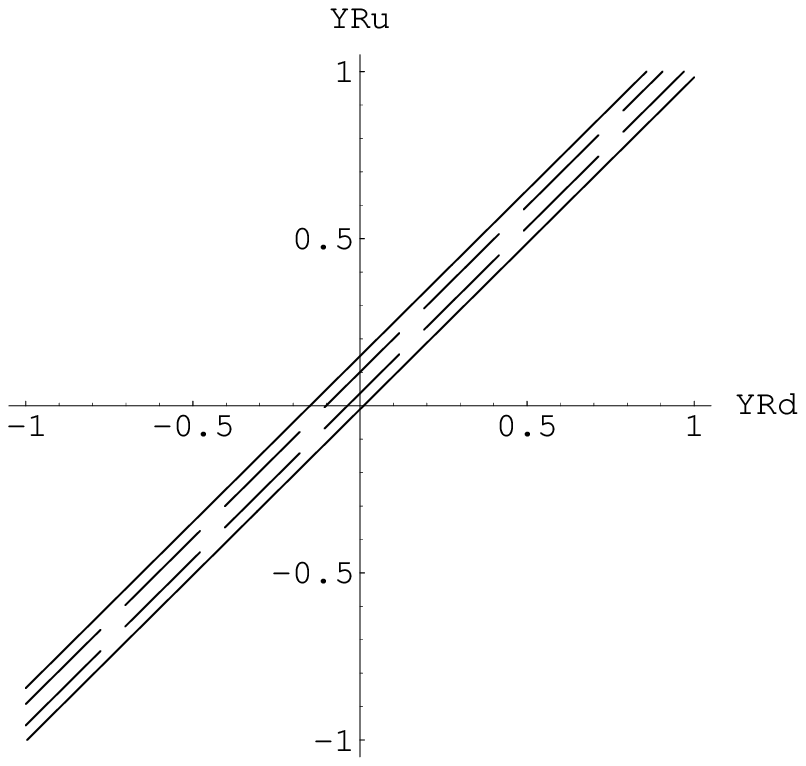}
    \caption{${\rm Re(\epsilon'/\epsilon)}$ as a function of $Y_{Rd}$ and
      $Y_{Ru}$ in the STC2 variant of Model~3. The $ 2\sigma$ and $5\sigma$
      bands are the same as in
      Fig.~\ref{fig:CPVB_fig_8}.}
    \label{fig:CPVB_fig_12}
  \end{center}
\end{figure}

\section*{Summary} 

We have reviewed how vacuum alignment in technicolor theories causes
spontaneous CP violation, and we described a possible natural solution to the
quarks' strong CP problem --- one which has {\em no axion} and {\em no
  massless up quark}. In these theories, flavor mixing and CP violation
appears in the standard weak interactions, as well as in new four-fermion ETC
and TC2 interactions. We explored the compatibility of these new effects with
current measurements, especially of $\stwobeta_{\rm eff}$. In contrast with
previous work~\cite{Lane:2002wv,Burdman:2000in}, we did not limit ourselves
to alignment models with mixing and CP violation solely in the left handed
quark sector, i.e., in the $D_L$ alignment matrix.

Mixing and CP violation in the $K^0$ system constrained the ETC gauge boson
masses to be so large that they do not contribute appreciably to any
$B$-meson decays or mixing. Therefore, we focused on TC2 interactions in
their standard (STC2) and flavor-universal (FUTC2) variants, working with
$\suone$ and $\uone$ couplings $\alpha_t(V_8)$ and $\alpha_t(Z')$ chosen
large enough to avoid their fine tuning. For each variant, we considered
three models of the quark mass matrices with $\arg\det(M_q) = 0$, designed to
give a fairly realistic CKM matrix, $V = U_L^\dagg D_L$, and various amounts
of mixing in the $D_R$ alignment matrix.

We found that $B_d$-mixing constraints require $M_{V_8,Z'} \simge
5$--$25\,\tev$ even in models with a non-block-diagonal $D_R$. These bounds
are as higher than those found in Ref.~\cite{Burdman:2000in} because of the
larger value of $|V_{td}|$ used here. They are considerably higher than the
$M_{Z'} \simge 1\,\tev$ estimated in Ref.~\cite{Simmons:2001va}. The
principal reason for that disagreement is our insistence on using a large
$\uone$ coupling $\alpha_q(Z')$ to avoid fine-tuning $\alpha_q(V_8)$. But, we
cannot win because the larger TC2 boson masses also require fairly severe fine
tuning, at the level of 3\%--0.05\%. Future measurements of $\Delta M_{B_s}$
may cause the TC2 gauge boson mass limit to increase further, as we found
that models with $M_{Z'} = M_{V_8} \cong 5\ \tev$ predict values for $\Delta
M_{B_s}\ {\rm and}\ x_s$ up to 15 times greater than the current experimental
bound.

Employing a minimal renormalization scheme, we calculated the effect of TC2
interactions on $\BD \to XK_S$ decays. We found that both variants of TC2 can
predict a $\stwobeta_{\rm eff}$ discrepancy among $\BD \to XK_S$ modes.
However, this discrepancy is directly related to the magnitudes and phase
difference between $D_{Rbq}$ and $D_{Lbq}$ matrix elements, and thus is
possible {\em only} for models with quark mass matrices with
non-block-diagonal $D_R$ (Models~2 and~3). In these models, we found that
moderate $Y_{Rd}, Y_{Ru}$ hypercharges are generally sufficient to achieve
discrepancies consistent with the current experimental values.

The contributions from TC2 to ${\rm Re}(\epsilon'/\epsilon)$ were also
calculated and found to be significant, even for models with block-diagonal
$D_R$. To accommodate the experimental value of ${\rm Re(\epsilon'/\epsilon)}$
in the alignment models considered, we must have $Y_{Rd} \cong Y_{Ru}$. This
further restricts the range of $\stwobeta$ values an individual alignment
model can generate, especially in the FUTC2 variant, as can be seen from
Figs.~\ref{fig:CPVB_fig_1}--\ref{fig:CPVB_fig_7}.

To sum up, new sources of flavor mixing and CP violation from TC2
interactions can be compatible with all constraints and still yield a
discrepancy in the observable $\stwobeta_{\rm eff}$. However, to accomplish
this fit, the $Z'$ and $V_8$ masses must be rather large so that TC2
interactions are fine-tuned at about the percent level. Somewhat
surprisingly, TC2 effects also tend to produce large values for ${\rm
  Re}(\epsilon'/\epsilon)$ and $\Delta M_{B_s}$. We may have to wait till the
end of this decade before we know the value of the latter.

\section*{Acknowledgements}

We are especially indebted to Gustavo Burdman for tutoring K.L. in
$B$-physics and early calculations of ${\rm Re}(\epsilon'/\epsilon)$ and to
Tongu\c c Rador and Estia Eichten for creating the suite of programs used for
quark vacuum alignment. Our thanks also go to Sekhar Chivukula and Yuval
Grossman for several helpful discussions.

\vfil\eject

\section*{Appendix A:  ETC Gauge Boson Mass Scales}

To set the ETC strengths $\Lambda^{TT},\Lambda^{Tq},\Lambda^{qq}$ in $\CH'$
of Eq.~(\ref{eq:Hetc}), we are assuming a TC2 model containing $N$ identical
electroweak doublets of technifermions. The technipion decay constant (which
sets the technicolor energy scale) is then $F_T = F_\pi/\sqrt{N}$, where
$F_\pi = 246\,\gev$ is the fundamental weak scale. We estimate the ETC masses
in $\CH'$ by the rule stated in Sect.~3: The ETC scale $\Metc/\getc$ in a
term involving weak eigenstates of the form $\ol q^{\, \prime}_i q'_j \ol
q^{\, \prime}_j q'_i$ or $\ol q^{\, \prime}_i q'_i \ol q^{\, \prime}_j q'_j$
(for $q'_i = u'_i$ or $d^{\, \prime}_i$) is approximately the same as the
scale that generates the $\ol q^{\, \prime}_{Ri} q'_{Lj}$ mass term,
$(\CM_q)_{ij}$.

The ETC gauge boson mass $\Metc(q)$ giving rise to a quark mass
$m_q(\Metc)$ --- an element or eigenvalue of $\CM_q$ --- is defined
by~\cite{Lane:2002wv}
\be\label{eq:qmass}
m_q(\Metc) \simeq \frac{g^2_{ETC}}{M^2_{ETC}(q)} \condetc \,.
\ee
Here, the quark mass and the technifermion bilinear condensate, $\condetc$,
are renormalized at the scale $\Metc(q)$. The condensate is related to the
one renormalized at the technicolor scale $\Ltc \simeq F_T$ by
\be\label{eq:condrenorm}
\condetc = \condtc \, \exp\left(\int_{\Ltc}^{\Metc(q)} \, \frac{d\mu}{\mu}
  \, \gamma_m(\mu) \right) \,.
\ee
Scaling $\condtc$ from QCD, we expect
\be\label{eq:ctc}
\condtc \equiv \Delta_T \simeq 4 \pi F^3_T = 4\pi F^3_\pi/N^{3/2} \,.
\ee
The anomalous dimension $\gamma_m$ of the operator $\ol T T$ is given in
perturbation theory by
\be\label{eq:gmma}
\gamma_m(\mu) = \frac{3 C_2(R)}{2 \pi}\atc(\mu) + O(\atc^2) \,,
\ee
where $C_2(R)$ is the quadratic Casimir of the technifermion $\sutc$
representation $R$. For the fundamental representation of $\sutc$ to which we
assume our technifermions $T$ belong, it is $C_2(\Ntc) = (\Ntc^2-1)/2\Ntc$.
In a walking technicolor theory, however, the coupling $\atc(\mu)$ decreases
very slowly from its critical chiral symmetry breaking value at $\Ltc$, and
$\gamma_m(\mu) \simeq 1$ for $\Ltc \simle \mu \simle \Metc$.

\begin{figure}[!h]
 \includegraphics[width=4.0in,height=6.0in,angle=90]{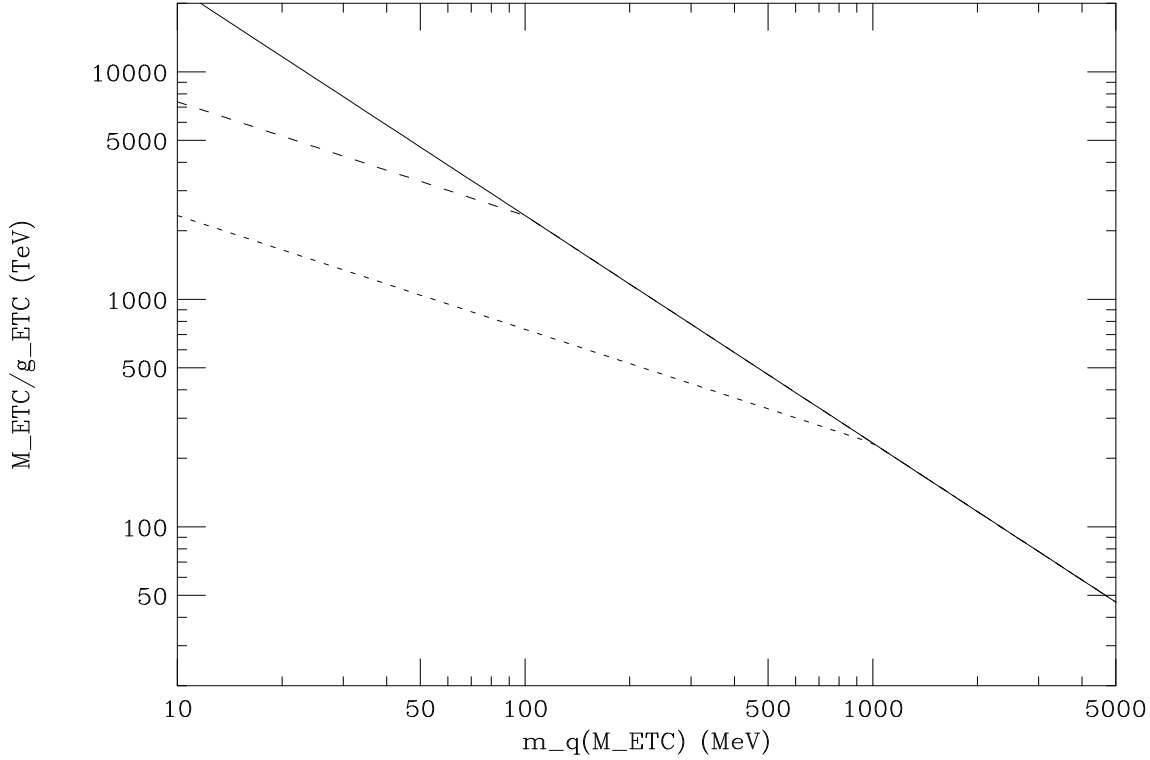}
 \caption{Extended technicolor scale $\Metc/\getc$ as a function of quark mass
   $m_q$ renormalized at $\Metc$ for $\kappa = 1$ (solid curve), $10$
   (dashed), and $100$ (solid).
    \label{fig:CPVB_fig_14}}
\end{figure}

An accurate evaluation of the condensate enhancement integral in
Eq.~(\ref{eq:condrenorm}) requires detailed specification of the technicolor
model and knowledge of the $\beta(\atc)$-function for large
coupling.\footnote{See Ref.~\cite{Lane:1991qh} for an attempt to calculate this
integral in a walking technicolor model.} Lacking this, we estimate the
enhancement by assuming that
\bea\label{eq:gmm}
\gamma_m(\mu) &= \left\{\ba{ll} 1 & {\rm for} \,\,\,\, \Ltc < \mu <
  \CM_{ETC}/\kappa \\
0 & {\rm for} \,\,\,\, \mu > \CM_{ETC}/\kappa
\ea \right.
\eea
Here, $\CM_{ETC}$ is the largest ETC scale, i.e., the one generating the
smallest term in the quark mass matrix for $\kappa =1$. The parameter $\kappa
> 1$ parameterizes the departure from the strict walking limit (which we
characterize by $\gamma_m = 1$ up to $\CM_{ETC}/\kappa$). Then, using
Eqs.~(\ref{eq:qmass},\ref{eq:condrenorm}), we obtain
\bea\label{eq:metc}
\frac{\Metc(q)}{\getc} &= \left\{\ba{ll} \frac{\sqrt{64\pi^3\alpha_{ETC}}\,
    F^2_\pi}{N m_q}  & {\rm if} \,\,\,\, \Metc(q) <
  \CM_{ETC}/\kappa \\ \\
  \sqrt{\frac{4\pi \CM_{ETC} F^2_\pi}{\kappa N m_q}} & {\rm if}
  \,\,\,\,
  \Metc(q) > \CM_{ETC}/\kappa
\ea \right.
\eea
To evaluate this, we take $\alpha_{ETC} = 3/4$, a moderately strong value as
would be expected in walking technicolor, $N = 10$, a typical number of
doublets in TC2 models with topcolor breaking (see, e.g.,
Ref.~\cite{Lane:1996ua}). Then, taking the smallest quark mass at the ETC
scale to be $10\,\mev$, we find $\CM_{ETC} = 7.17\times 10^4\,\tev$. The
resulting estimates of $\Metc/\getc$ are plotted in
Fig.~\ref{fig:CPVB_fig_14} for $\kappa = 1, 10$, and 100. They run from
$\Metc/\getc = 47\,\tev$ for $m_q = 5\,\gev$ to $2.34/\sqrt{\kappa}\times
10^4\,\tev/$ for $m_q = 10\,\mev$. Very similar results are obtained for
$\alpha_{ETC} = 1/2$ and $N=8$: $\Metc/\getc = 48\,\tev$ for $m_q = 5\,\gev$
to $2.38/\sqrt{\kappa}\times 10^4\,\tev/$ for $m_q = 10\,\mev$.

\section*{Appendix B: Three Mass Models and their Alignment Matrices}

In Sect.~2 we described a scenario for solving the strong CP problem of QCD.
Briefly, it was based on the natural appearance of vacuum-aligning phases in
the technicolor sector which are rational multiples of $\pi$ {\em and} the
assumption that ETC interactions map these phases onto the primordial quark
mass matrix $\CM_q$ of Eq.~(\ref{eq:primordial}) in such a way that
\be\label{eq:argdetMq}
\arg\det(\CM_q) \equiv \arg\det(\CM_u) + \arg\det(\CM_d) = 0 \,.
\ee
Since $\CM_q$ in this case is brought to real, positive, diagonal form $M_q$
by unimodular matrices $Q_{L,R} = (U,D)_{L,R}$, then $\arg\det(M_q) = 0$
also. In the absence of an explicit ETC model, we cannot construct $\CM_q$
from ``first principles''. In this appendix, therefore, we write down mass
matrices with rational phases satisfying Eq.~(\ref{eq:argdetMq}), carry out
vacuum alignment in the quark sector by minimizing the energy $E_{Tq}(Q = Q_L
Q_R^\dagg) \cong - {\rm Tr}(\CM_q \, Q + {\rm h.c.})\Delta_q(\Metc)$, and
thus determine the aligning matrices $Q_{L,R}$, and the CKM matrix $V=
U_L^\dagg D_L$ with its unphysical phases removed.\footnote{The programs to
  carry out quark-sector vacuum alignment were developed by Tongu\c c Rador
  in 2000.} We remind the reader that these $\CM_q$ have not been
``fine-tuned'' to give $|V_{ui d_j}|$ in complete accord with values found in
Ref.~\cite{Eidelman:2004wy}

\subsection*{Mass Model 1: Block-Diagonal $D_R$}

As we discussed in Sect.~2, there is an argument, albeit one we believe is
questionable, that $\CM_d$ must have a nearly triangular texture to suppress
$\BD$--$\ol \BD$ mixing induced by bottom pions. That texture is needed to
make $|D_{Rbd}| \ll |D_{Lbd}|$. This model has such a primordial $\CM_d$.
Then, because the very large TC2 contribution to $(\CM_u)_{tt}$ makes mixing
angles between $u,c$ and $t$ very small, all mixing between the heavy and
light generations in this model occurs in $D_L$. To create small $m_u$ and
$m_d$ while having a large CKM angle $\theta_{12} \simeq 0.2$, we take mass
matrices with a seesaw structure:
\bea\label{eq:pmqMM1}
\CM_u(\Metc) &=&\left(\ba{ccc} (0,0) & (200,0)& (0,0)\\
                               (16,2\pi/3) & (900,0) & (0,0)\\
                               (0,0) & (0,0) & (160000,0)\ea\right) \,;\nn\\\\
\CM_d(\Metc) &=&\left(\ba{ccc} (0,0) & (20, -\pi/3) & (0,0)\\
                               (22,0) & (100, 0) & (0,0)\\
                               (17,0) & (145, -\pi/3) & (3500,
                               -\pi/3)\ea\right)\,.\nn
\eea
The notation is $(|(\CM_q)_{ij}|, \arg((\CM_q)_{ij}))$, with masses in $\mev$,
and these matrices are renormalized at $\Metc \sim 10^3\,\tev$. The quark
mass eigenvalues extracted from $\CM_q$ are:
\bea\label{eq:masses}
m_u &=& 3.47\ts, \ts\ts\ts m_c = 922\ts, \ts\ts\ts m_t = 160000\,; \nn\\
m_d &=& 4.22\ts, \ts\ts\ts m_2 = 104\ts, \ts\ts\ts m_b = 3503\,.
\eea
Quark masses at the EW scale will be enhanced by QCD and, possibly, TC2
renormalizations which we will not carry out. For the calculations in
Sects.~4--7, we took quark masses from Ref.~\cite{Eidelman:2004wy}.

The $U(3)$ matrices $U = U_{L}U^{\dag}_{R}$ and $D = D_LD^{\dag}_R$
obtained by minimizing the vacuum energy $E_{Tq}$ are:
\bea\label{eq:UMM1}
U &=& \left(\ba{ccc} (0.972, \pi/3) & (0.233, -2\pi/3) & (0,0)\\
                        (0.233,0 ) & (.972, 0) & (0,0)\\
                        (0,0) & (0,0) & (1.000, 0)\ea\right) \ts;
                      \nn\\\nn\\
D  &=&\left(\ba{ccc} (0.922, -2\pi/3) & (0.387, 0) & (0.0047,0.014\pi )\\
                        (0.387, \pi/3) & (0.921, 0) & (0.0402,\pi/3)\\
                        (0.0141, -0.755\pi) & (0.0379, -0.986\pi) & (0.999, \pi/3)
                        \ea\right) \ts; \nn  
\eea
The minimization routine we use must determine the $6\times 6$ block-diagonal
$Q = (U,D)$ all at once because it, not $U$ and $D$ separately, is
unimodular. It is obvious that large elements of $U$ and $D$ have phases
which are rational multiples of $\pi$, reflecting those in $\CM_u$ and
$\CM_d$. But the program has a little difficulty determining precisely the
phases for small matrix elements. For example, the phase $0.014\pi$ of
$D_{db}$ is probably zero. We do not believe there are large errors in any of
the phases in these matrices. Therefore, the observable combinations of
phases in $V = U_L^\dagg D_L$ and $Q_{L,R}$ below should be well-determined.

By Nuyts' theorem~\cite{Nuyts:1971az}, the mass matrix $\CM_q Q$ which
minimizes the vacuum energy is diagonalized by the single block-diagonal
$SU(6)$ matrix $Q_R = (U,D)_R$. With $Q$ and $Q_R$ determined, we obtain $Q_L
= Q Q_R$ and $M_q = Q_R^\dagg \CM_q Q_L$. We then construct $V$ and remove
its five unobservable phases to put it in the standard form,
Eq.~(\ref{eq:CKMmat}). This leaves $2\times 6 - 1 - 5 = 6$ independent phases
in the $Q_L$. To maintain the vacuum's alignment, these five $q_{Li}$ phase
changes must be accompanied by the same transformations on the $q_{Ri}$. No
further quark phase changes are permissible, leaving $Q_R$ with
11~independent phases. One of these is the overall $T_{R3}$ angle,
$\arg\det(D_R) = -\arg\det(D_R)$. Only the remaining 10~$Q_R$-phases, five
each in $U_R$ and $D_R$, appear in the right-handed flavor ETC and TC2
currents and are, in principle, measurable.

The phase-adjusted CKM matrix is:
\bea\label{eq:VMM1}
V &=& \left(\ba{ccc} (0.976, 0) & (0.216, 0)  & (0.0045, -0.977)\\ 
                  (0.216, 3.142) & (0.976, 0) & (0.0415, 0)\\
                  (0.0075, -0.516) & (0.0410, 3.161) & (0.999, 0)
                  \ea\right) \ts. \nn\\
\eea
The angles $\theta_{ij}$ and phase $\delta_{13}$ corresponding to this matrix
are:
\be\label{eq:angsMM1}
\theta_{12} = 0.218 \ts, \ts\ts\ts \theta_{23} = 0.0415 \ts, \ts\ts\ts
\theta_{13} = 0.00455 \ts, \ts\ts\ts \delta_{13} = 0.977 \ts.
\ee
The magnitudes of the $V_{u_i d_j}$ are in fair agreement with those in
Ref.~\cite{Eidelman:2004wy}. The phase-adjusted $Q_{L,R}$ matrices are (up to
terms of $\CO(10^{-5}))$:
\bea\label{eq:UDMM1}
U_L &=&\left(\ba{ccc} (1.000, 0.291) & (0.017, -1.867) & (0,0)\\
                        (0.017,-0.756 ) & (1.000, 0.227) & (0,0)\\
                        (0,0) & (0,0) & (1.000, 0.225)\ea\right) \ts;
                      \nn\\\nn\\
D_L &=&\left(\ba{ccc} (0.978, 0.294) & (0.207, 0.224) & (0.0049, -0.821)\\
                        (0.207, 3.435) & (0.977, 0.224) & (0.041,0.225)\\
                        (0.0074, -0.291) & (0.041, 3.387) & (0.999, 0.225)
                        \ea\right) \ts; \nn \\ \\
U_R &=&\left(\ba{ccc} (0.976, -0.756) & (0.217, 0.227) & (0,0)\\
                        (0.216, 2.385) & (0.976, 0.227) & (0,0)\\
                        (0,0) & (0,0) & (1, 0.225)\ea\right) \ts; \nn\\ 
D_R &=&\left(\ba{ccc} (0.982, 2.388) & (0.188, -0.824) & (2.4\times 10^{-4}, -0.822)\\ 
                        (0.188, 0.294) & (0.982,0.223) & (0.0012, 0.203)\\ 
                        (0,0) & (0.0012, 2.34) & (1.000, -0.822)
                        \ea\right) \ts.\nn
\eea
By design, $D_R$ is very nearly block diagonal.

\subsection*{Mass Model 2: Nontrivial $D_R$}

This model has nonzero heavy-light generational mixing in $D_R$. Elements are
again of the form $(|(\CM_q)_{ij}|, \arg((\CM_q)_{ij}))$:
\bea\label{eq:pmqMM2}
\CM_u(\Metc) &=&\left(\ba{ccc} (0,0) & (200,0)& (0,0)\\
                               (16,2\pi/3) & (900,0) & (0,0)\\
                               (0,0) & (0,0) & (160000,0)\ea\right)\,;\nn\\\\
\CM_d(\Metc) &=&\left(\ba{ccc} (0,0) & (20, -\pi/3) & (0,0)\\
                               (22,0) & (100, 0) & (140, -\pi/3)\\
                               (17,0) & (100, -\pi/3) & (3500,
                               -\pi/3)\ea\right)\,.\nn
\eea
The primordial mass matrices in Model~2 are similar to those of Model~1,
except that the off-diagonal elements $(\CM_d)_{bs}$ and $(\CM_{d})_{sb}$ are
comparable. For this reason, the physical masses of Model~2 are practically
identical to those of Model~1, but we will not obtain a block-diagonal $D_R$.
The matrices $U, D$ minimizing the vacuum energy are:
\bea\label{eq:UMM2}
U &=& \left(\ba{ccc} (0.972, \pi/3) & (0.233, -2\pi/3) & (0,0)\\
                        (0.233,0 ) & (.972, 0) & (0,0)\\
                        (0,0) & (0,0) & (1.000, 0)\ea\right) \ts;
                      \nn\\\nn\\
D  &=&\left(\ba{ccc} (0.921, -0.655\pi) & (0.388, 0) & (0.010, -0.996\pi )\\
                        (0.388, \pi/3) & (0.921, -0.011\pi) & (0.0334,0.727\pi)\\
                        (0.0097, -0.794\pi) & (0.0336, 0.566\pi) & (0.999, \pi/3)
                        \ea\right) \ts; \nn  
\eea

The phase-adjusted CKM matrix is:
\bea\label{eq:VMM2}
V &=& \left(\ba{ccc} (0.977, 0) & (0.214, 0)  & (0.0049, 5.259)\\ 
                  (0.214, 3.142) & (0.976, 0) & (0.0292, 0)\\
                  (0.0055, -0.826) & (0.0290, 3.172) & (1.000, 0)
                  \ea\right) \ts. \nn\\
\eea
The corresponding angles $\theta_{ij}$ and phase $\delta_{13}$ are:
\be\label{eq:angsMM2}
\theta_{12} = 0.216 \ts, \ts\ts\ts \theta_{23} = 0.0292 \ts, \ts\ts\ts
\theta_{13} = 0.00489 \ts, \ts\ts\ts \delta_{13} = 1.024 \ts.
\ee
The most important features of Model~2 are the size of the CKM element
$|V_{td}| = 0.0055$, $|V_{ts}| = 0.029$ compared to $|V_{td}| = 0.0075$,
$|V_{ts}| = 0.041$ in Model~1. The smaller $|V_{td}|$ is, the lighter their
$Z'$ and $V_8$ can be while still complying with the constraints from $\BD$
mixing. Lighter gauge bosons then lead to larger TC2 contributions to decay
and mixing processes. The CKM element $|V_{ts}|$ in Model 2 is only $\sim
70\%$ of the value in Ref.~\cite{Eidelman:2004wy}, which is known by the
unitarity relation $|V_{ts}| \cong |V_{cb}| \cong 0.040$. The calculation of
$\Delta M_{B_s}$ is affected the most by this discrepancy as explained in
Section~7.2.

The phase-adjusted alignment matrices are: 
\bea\label{eq:UDMM2}
U_L &=&\left(\ba{ccc} (1.000, -1.842) & (0.019, -4.035) & (0,0)\\
                        (0.0169,-2.889) & (1.000, -1.941) & (0,0)\\
                        (0,0) & (0,0) & (1.000, -1.909)\ea\right) \ts;
                      \nn\\\nn\\
D_L &=&\left(\ba{ccc} (0.979, -1.839) & (0.205, -1.908) & (0.0051, 3.328)\\
                        (0.205, 1.267) & (0.978, -1.944) & (0.029,-1.943)\\
                        (0.0055, 3.548) & (0.0290, 1.263) & (0.999, -1.909)
                        \ea\right) \ts; \nn \\ \\
U_R &=&\left(\ba{ccc} (0.976, -2.889) & (0.217, -1.941) & (0,0)\\
                        (0.217, 0.252) & (0.976, -1.941) & (0,0)\\
                        (0,0) & (0,0) & (1, -1.909)\ea\right) \ts; \nn\\ \nn\\
D_R &=&\left(\ba{ccc} (0.981, 0.220) & (0.192, 3.292) & (1.6\times 10^{-4}, 3.293)\\ 
                        (0.191, -1.839) & (0.981,-1.908) & (0.040, 3.344)\\ 
                        (0.0077, 1.303) & (0.0397,1.215) &
                        (0.999, 3.327)
                        \ea\right) \nn \ts.
\eea
Note that $|D_{Rbq}|$ is actually larger than $|D_{Lbq}|$ for $q=d,s$.

\subsection*{Mass Model 3: Nontrivial $D_R$ with $|D_{Rbq}|\simeq |D_{Lbq}|$}

This model also has a nontrivial heavy-light generational mixing in $D_R$, but
we start from a more symmetric $\CM_d$ than in Mass Model~2:
\bea\label{eq:pmqMM3}
\CM_u(\Metc) &=&\left(\ba{ccc} (0,0) & (200,0)& (0,0)\\
                               (16,2\pi/3) & (900,0) & (0,0)\\
                               (0,0) & (0,0) & (160000,0)\ea\right)\,;\nn\\\\
\CM_d(\Metc) &=&\left(\ba{ccc} (0,0) & (20, -\pi/3) & (0,0)\\
                               (22,0) & (100, 0) & (100, -\pi/3)\\
                               (17,0) & (100, -\pi/3) & (3500,
                               -\pi/3)\ea\right)\,.\nn
\eea
The physical masses at $M_{ETC}$ are again almost identical to those of
Model~1. The minimizing matrices $U, D$ are:
\bea\label{eq:UMM3}
U &=& \left(\ba{ccc} (0.972, \pi/3) & (0.233, -2\pi/3) & (0,0)\\
                        (0.233,0 ) & (.972, 0) & (0,0)\\
                        (0,0) & (0,0) & (1.000, 0)\ea\right) \ts;
                      \nn\\\nn\\
D  &=&\left(\ba{ccc} (0.922, -0.659\pi) & (0.388, 0) & (0.0061, -0.993\pi )\\
                        (0.388, \pi/3) & (0.921, -0.008\pi) & (0.0273,0.640\pi)\\
                        (0.0096, -0.796\pi) & (0.0263, 0.659\pi) & (0.999, \pi/3)
                        \ea\right) \ts; \nn  
\eea

The CKM matrix in standard form is: 
\bea\label{eq:VMM3}
V &=& \left(\ba{ccc} (0.977, 0) & (0.215, 0)  & (0.0048, -1.023)\\ 
                  (0.215, 3.142) & (0.976, 0) & (0.0290, 0)\\
                  (0.0055, -0.816) & (0.0289, 3.172) & (1.000, 0)
                  \ea\right) \ts. \nn\\
\eea
Its angles $\theta_{ij}$ and phase $\delta_{13}$ are
\be\label{eq:angsMM3}
\theta_{12} = 0.216 \ts, \ts\ts\ts \theta_{23} = 0.0290 \ts, \ts\ts\ts
\theta_{13} = 0.00482 \ts, \ts\ts\ts \delta_{13} = 1.023 \ts.
\ee
Like Mass Model~2, this model has a small CKM $V_{td}$ element. We therefore
expect lower $Z'$ and $V_8$ mass bounds than in Model~1. That is indeed the
case, but they are somewhat larger than in Model~2. This demonstrates the
difficulty of obtaining model-independent lower bounds on $M_{Z',V_8}$ from
the $\BD$-mixing constraint with non-block diagonal $D_R$ --- a situation
already emphasized by Simmons~\cite{Simmons:2001va}.

The phase-adjusted $Q_{L,R}$ are:
\bea\label{eq:UDMM3}
U_L &=&\left(\ba{ccc} (1.000, -1.891) & (0.017, -4.074) & (0,0)\\
                        (0.017,-2.939) & (1.000, -1.980) & (0,0)\\
                        (0,0) & (0,0) & (1.000, -1.958)\ea\right) \ts;
                      \nn\\\nn\\
D_L &=&\left(\ba{ccc} (0.979, -1.889) & (0.205, -1.958) & (0.0050, 3.278)\\
                        (0.205, 1.228) & (0.978, -1.983) & (0.0289,-1.983)\\
                        (0.0055, 3.509) & (0.0289, 1.214) & (1.000, -1.958)
                        \ea\right) \ts; \nn \\\nn \\
U_R &=&\left(\ba{ccc} (0.976, -2.939) & (0.217, -1.980) & (0,0)\\
                        (0.217, 0.203) & (0.976, -1.980) & (0,0)\\
                        (0,0) & (0,0) & (1.000, -1.958)\ea\right) \ts; \nn\\ \nn\\
D_R &=&\left(\ba{ccc} (0.982, 0.181) & (0.190, 3.253) & (1.6\times 10^{-4}, 3.253)\\ 
                        (0.190, -1.888) & (0.981,-1.958) & (0.0290, 3.302)\\ 
                        (0.0054, 1.254) & (0.0285,1.158) & (1.000, 3.278)
                        \ea\right) \ts.
\eea
The effect of the more symmetric $\CM_d$ is seen in the $(bd)$ and $(bs)$
elements of $D_L$ and $D_R.$

\vfil\eject

\bibliography{CPVB}

\providecommand{\href}[2]{#2}\begingroup\raggedright\begin{thebibliography}{10}

\bibitem{Browder:2003ii}
T.~E. Browder, ``Results on the CKM angle phi(1)(beta),''
  \href{http://xxx.lanl.gov/abs/hep-ex/0312024}{ hep-ex/0312024}.

\bibitem{Aubert:2004dy}
{\bf BABAR} Collaboration, B.~Aubert {\em et.~al.}, ``Measurements of CP
  asymmetries in the decay B --> Phi K,''
  \href{http://xxx.lanl.gov/abs/hep-ex/0408072}{ hep-ex/0408072}.

\bibitem{Abe:2004xp}
{\bf BELLE} Collaboration, K.~Abe {\em et.~al.}, ``New measurements of
  time-dependent CP-violating asymmetries in b --> s transitions at Belle,''
  \href{http://xxx.lanl.gov/abs/hep-ex/0409049}{ hep-ex/0409049}.

\bibitem{Aubert:2003bq}
{\bf BABAR} Collaboration, B.~Aubert {\em et.~al.}, ``Measurements of
  CP-violating asymmetries and branching fractions in $B$ meson decays to
  $\eta' K$,'' {\em Phys. Rev. Lett.} {\bf 91} (2003) 161801,
  \href{http://xxx.lanl.gov/abs/hep-ex/0303046}{ hep-ex/0303046}.

\bibitem{Aubert:2004xf}
{\bf BABAR} Collaboration, B.~Aubert {\em et.~al.}, ``Measurements of CP
  violating asymmetries in $B^0 \to K^0_S \pi^0$ decays,''
  \href{http://xxx.lanl.gov/abs/hep-ex/0403001}{ hep-ex/0403001}.

\bibitem{Hill:2002ap}
C.~T. Hill and E.~H. Simmons, ``Strong dynamics and electroweak symmetry
  breaking,'' {\em Physics Reports} {\bf 381} (2003) 235--402,
  \href{http://xxx.lanl.gov/abs/hep-ph/0203079}{ hep-ph/0203079}.

\bibitem{Lane:2002wv}
K.~Lane, ``Two lectures on technicolor,''
  \href{http://xxx.lanl.gov/abs/hep-ph/0202255}{ hep-ph/0202255}. Lectures at
  l'Ecole de GIF, Annecy--le--Vieux, France, September 10--14, 2001.

\bibitem{Eichten:1980ah}
E.~Eichten and K.~Lane, ``Dynamical breaking of weak interaction symmetries,''
  {\em Phys. Lett.} {\bf B90} (1980) 125--130.

\bibitem{Holdom:1981rm}
B.~Holdom, ``Raising the sideways scale,'' {\em Phys. Rev.} {\bf D24} (1981)
  1441.

\bibitem{Appelquist:1986an}
T.~W. Appelquist, D.~Karabali, and L.~C.~R. Wijewardhana, ``Chiral hierarchies
  and the flavor changing neutral current problem in technicolor,'' {\em Phys.
  Rev. Lett.} {\bf 57} (1986) 957.

\bibitem{Yamawaki:1986zg}
K.~Yamawaki, M.~Bando, and K.-i. Matumoto, ``Scale invariant technicolor model
  and a technidilaton,'' {\em Phys. Rev. Lett.} {\bf 56} (1986) 1335.

\bibitem{Akiba:1986rr}
T.~Akiba and T.~Yanagida, ``Hierarchic chiral condensate,'' {\em Phys. Lett.}
  {\bf B169} (1986) 432.

\bibitem{Hill:1995hp}
C.~T. Hill, ``Topcolor assisted technicolor,'' {\em Phys. Lett.} {\bf B345}
  (1995) 483--489, \href{http://xxx.lanl.gov/abs/hep-ph/9411426}{
  hep-ph/9411426}.

\bibitem{Chivukula:1998vd}
R.~S. Chivukula and H.~Georgi, ``Effective field theory of vacuum tilting,''
  {\em Phys. Rev.} {\bf D58} (1998) 115009,
  \href{http://xxx.lanl.gov/abs/hep-ph/9806289}{ hep-ph/9806289}.

\bibitem{Chivukula:1996cc}
R.~S. Chivukula and J.~Terning, ``Precision electroweak constraints on
  topcolor-assisted technicolor,'' {\em Phys. Lett.} {\bf B385} (1996)
  209--217, \href{http://xxx.lanl.gov/abs/hep-ph/9606233}{ hep-ph/9606233}.

\bibitem{Lane:1996ua}
K.~Lane, ``Symmetry breaking and generational mixing in topcolor-assisted
  technicolor,'' {\em Phys. Rev.} {\bf D54} (1996) 2204--2212,
  \href{http://xxx.lanl.gov/abs/hep-ph/9602221}{ hep-ph/9602221}.

\bibitem{Chivukula:1996yr}
R.~S. Chivukula, A.~G. Cohen, and E.~H. Simmons, ``New strong interactions at
  the Tevatron?,'' {\em Phys. Lett.} {\bf B380} (1996) 92--98,
  \href{http://xxx.lanl.gov/abs/hep-ph/9603311}{ hep-ph/9603311}.

\bibitem{Popovic:1998vb}
M.~B. Popovic and E.~H. Simmons, ``A heavy top quark from flavor-universal
  colorons,'' {\em Phys. Rev.} {\bf D58} (1998) 095007,
  \href{http://xxx.lanl.gov/abs/hep-ph/9806287}{ hep-ph/9806287}.

\bibitem{Lane:2001ar}
K.~Lane, ``$K^0$--$\bar K^0$ and $B^0$--$\bar B^0$ constraints on
  technicolor,'' \href{http://xxx.lanl.gov/abs/hep-ph/0106279}{
  hep-ph/0106279}. Invited talk at Les Rencontres de Physique de la Vall\'ee
  d'Aoste, La Thuile, Italy, March 4-10, 2001.

\bibitem{Lane:2001rv}
K.~Lane, ``Strong and weak CP violation in technicolor,''
  \href{http://xxx.lanl.gov/abs/hep-ph/0106328}{ hep-ph/0106328}. Invited talk
  at the Eighth International Symposium on Particles, Strings and
  Cosmology---PASCOS 2001, University of North Carolina, Chapel Hill, NC, April
  10--15, 2001.

\bibitem{Burdman:2000in}
G.~Burdman, K.~D. Lane, and T.~Rador, ``$\bar B$--$B$ mixing constrains
  topcolor-assisted technicolor,'' {\em Phys. Lett.} {\bf B514} (2001) 41--46,
  \href{http://xxx.lanl.gov/abs/hep-ph/0012073}{ hep-ph/0012073}.

\bibitem{Simmons:2001va}
E.~H. Simmons, ``Limitations of B meson mixing bounds on technicolor
  theories,'' {\em Phys. Lett.} {\bf B526} (2002) 365--369,
  \href{http://xxx.lanl.gov/abs/hep-ph/0111032}{ hep-ph/0111032}.

\bibitem{Burdman:2003nt}
G.~Burdman, ``Flavor violation in warped extra dimensions and CP asymmetries in
  B decays,'' \href{http://xxx.lanl.gov/abs/hep-ph/0310144}{ hep-ph/0310144}.

\bibitem{Dashen:1971et}
R.~F. Dashen, ``Some features of chiral symmetry breaking,'' {\em Phys. Rev.}
  {\bf D3} (1971) 1879--1889.

\bibitem{Eichten:1980du}
E.~Eichten, K.~Lane, and J.~Preskill, ``{CP} violation without elementary
  scalar fields,'' {\em Phys. Rev. Lett.} {\bf 45} (1980) 225.

\bibitem{Lane:1981je}
K.~Lane, ``CP nonconservation in dynamically broken gauge theories,'' {\em
  Phys. Scripta} {\bf 23} (1981) 1005.

\bibitem{Lane:2000es}
K.~Lane, T.~Rador, and E.~Eichten, ``Vacuum alignment in technicolor theories.
  I: The technifermion sector,'' {\em Phys. Rev.} {\bf D62} (2000) 015005,
  \href{http://xxx.lanl.gov/abs/hep-ph/0001056}{ hep-ph/0001056}.

\bibitem{Peskin:1990zt}
M.~E. Peskin and T.~Takeuchi, ``A new constraint on a strongly interacting
  Higgs sector,'' {\em Phys. Rev. Lett.} {\bf 65} (1990) 964--967.

\bibitem{Holdom:1990tc}
B.~Holdom and J.~Terning, ``Large corrections to electroweak parameters in
  technicolor theories,'' {\em Phys. Lett.} {\bf B247} (1990) 88--92.

\bibitem{Golden:1991ig}
M.~Golden and L.~Randall, ``Radiative corrections to electroweak parameters in
  technicolor theories,'' {\em Nucl. Phys.} {\bf B361} (1991) 3--23.

\bibitem{Peccei:1996ax}
R.~D. Peccei, ``QCD, strong CP and axions,'' {\em J. Korean Phys. Soc.} {\bf
  29} (1996) S199--S208, \href{http://xxx.lanl.gov/abs/hep-ph/9606475}{
  hep-ph/9606475}.

\bibitem{Nuyts:1971az}
J.~Nuyts, ``Is CP-invariance violation caused by an SU(3) singlet?,'' {\em
  Phys. Rev. Lett.} {\bf 26} (1971) 1604--1605.

\bibitem{Harari:1986xf}
H.~Harari and M.~Leurer, ``Recommending a standard choice of Cabibbo angles and
  KM phases for any number of generations,'' {\em Phys. Lett.} {\bf B181}
  (1986) 123.

\bibitem{Eidelman:2004wy}
{\bf Particle Data Group} Collaboration, S.~Eidelman {\em et.~al.}, ``Review of
  particle physics,'' {\em Phys. Lett.} {\bf B592} (2004) 1.

\bibitem{Buchalla:1996dp}
G.~Buchalla, G.~Burdman, C.~T. Hill, and D.~Kominis, ``GIM violation and new
  dynamics of the third generation,'' {\em Phys. Rev.} {\bf D53} (1996)
  5185--5200, \href{http://xxx.lanl.gov/abs/hep-ph/9510376}{ hep-ph/9510376}.

\bibitem{Kominis:1995fj}
D.~Kominis, ``Flavor changing neutral current constraints in topcolor assisted
  technicolor,'' {\em Phys. Lett.} {\bf B358} (1995) 312--317,
  \href{http://xxx.lanl.gov/abs/hep-ph/9506305}{ hep-ph/9506305}.

\bibitem{Lane:1995gw}
K.~Lane and E.~Eichten, ``Natural topcolor assisted technicolor,'' {\em Phys.
  Lett.} {\bf B352} (1995) 382--387,
  \href{http://xxx.lanl.gov/abs/hep-ph/9503433}{ hep-ph/9503433}.

\bibitem{Lane:1998qi}
K.~Lane, ``A new model of topcolor-assisted technicolor,'' {\em Phys. Lett.}
  {\bf B433} (1998) 96--101, \href{http://xxx.lanl.gov/abs/hep-ph/9805254}{
  hep-ph/9805254}.

\bibitem{Buras:2001pn}
A.~J. Buras, ``Flavor dynamics: CP violation and rare decays,''
  \href{http://xxx.lanl.gov/abs/hep-ph/0101336}{ hep-ph/0101336}.

\bibitem{Buras:1998fb}
A.~J. Buras and R.~Fleischer, ``Quark mixing, CP violation and rare decays
  after the top quark discovery,'' {\em Adv. Ser. Direct. High Energy Phys.}
  {\bf 15} (1998) 65--238, \href{http://xxx.lanl.gov/abs/hep-ph/9704376}{
  hep-ph/9704376}.

\bibitem{Kramer:1994in}
G.~Kramer, W.~F. Palmer, and H.~Simma, ``CP violation and strong phases from
  penguins in B+- $\to$ P P and B+- $\to$ V P decays,'' {\em Z. Phys.} {\bf
  C66} (1995) 429--438, \href{http://xxx.lanl.gov/abs/hep-ph/9410406}{
  hep-ph/9410406}.

\bibitem{Grossman:1997gr}
Y.~Grossman, G.~Isidori, and M.~P. Worah, ``CP asymmetry in B/d --> Phi K(S):
  Standard model pollution,'' {\em Phys. Rev.} {\bf D58} (1998) 057504,
  \href{http://xxx.lanl.gov/abs/hep-ph/9708305}{ hep-ph/9708305}.

\bibitem{Buchalla:1996vs}
G.~Buchalla, A.~J. Buras, and M.~E. Lautenbacher, ``Weak decays beyond leading
  logarithms,'' {\em Rev. Mod. Phys.} {\bf 68} (1996) 1125--1144,
  \href{http://xxx.lanl.gov/abs/hep-ph/9512380}{ hep-ph/9512380}.

\bibitem{Ali:1998eb}
A.~Ali, G.~Kramer, and C.-D. Lu, ``Experimental tests of factorization in
  charmless nonleptonic two-body B decays,'' {\em Phys. Rev.} {\bf D58} (1998)
  094009, \href{http://xxx.lanl.gov/abs/hep-ph/9804363}{ hep-ph/9804363}.

\bibitem{Ali:1998nh}
A.~Ali and C.~Greub, ``An analysis of two-body nonleptonic B decays involving
  light mesons in the standard model,'' {\em Phys. Rev.} {\bf D57} (1998)
  2996--3016, \href{http://xxx.lanl.gov/abs/hep-ph/9707251}{ hep-ph/9707251}.

\bibitem{Fleischer:2002ys}
R.~Fleischer, ``CP violation in the B system and relations to $K \to \pi \nu
  \bar \nu$ decays,'' {\em Phys. Rept.} {\bf 370} (2002) 537--680,
  \href{http://xxx.lanl.gov/abs/hep-ph/0207108}{ hep-ph/0207108}.

\bibitem{Buras:2003zz}
A.~J. Buras and M.~Jamin, ``Epsilon'/epsilon at the NLO: 10 years later,'' {\em
  JHEP} {\bf 01} (2004) 048, \href{http://xxx.lanl.gov/abs/hep-ph/0306217}{
  hep-ph/0306217}.

\bibitem{Buras:2001au}
A.~J. Buras and J.-M. Gerard, ``What is the (epsilon'/epsilon)(exp) telling
  us?,'' {\em Phys. Lett.} {\bf B517} (2001) 129--134,
  \href{http://xxx.lanl.gov/abs/hep-ph/0106104}{ hep-ph/0106104}.

\bibitem{Lane:1991qh}
K.~D. Lane and M.~V. Ramana, ``Walking technicolor signatures at hadron
  colliders,'' {\em Phys. Rev.} {\bf D44} (1991) 2678--2700.

\end{thebibliography}\endgroup
\bibliographystyle{utcaps}
\end{document}